\documentclass[final,hidelinks,onefignum,onetabnum]{siamart251104}

\usepackage{multirow}
\usepackage{booktabs}
\usepackage{tikz}
\usepackage{wrapfig}
\usepackage{lipsum} 
\usepackage{amsmath}
\usepackage{mathtools}
\usepackage{pifont}
\usepackage{amssymb}
\usepackage{bbm}
\usepackage[numbers]{natbib} 
\usepackage{hyperref}

\usepackage{lipsum}
\usepackage{amsfonts}
\usepackage{graphicx}
\usepackage{epstopdf}
\usepackage{algorithmic}
\ifpdf
  \DeclareGraphicsExtensions{.eps,.pdf,.png,.jpg}
\else
  \DeclareGraphicsExtensions{.eps}
\fi

\newsiamremark{remark}{Remark}
\newsiamremark{hypothesis}{Hypothesis}
\crefname{hypothesis}{Hypothesis}{Hypotheses}
\newsiamthm{claim}{Claim}
\newsiamremark{fact}{Fact}
\crefname{fact}{Fact}{Facts}

\headers{Sequential Design Using Deep Gaussian Processes}{Y. Yang, D. Ming, and S. Guillas}

\title{Sequential Design for Simulators with Sharp Variations via Deep Gaussian Process Gradients} 

\author{Yiming Yang\thanks{Department of Statistical Science, University College London, London WC1E 6BT, UK
  (\email{zcahyy1@ucl.ac.uk,s.guillas@ucl.ac.uk}).}
\and Deyu Ming\thanks{School of Management,
University College London, London WC1E 6BT, UK
  (\email{deyu.ming.16@ucl.ac.uk}).}
\and Serge Guillas\textsuperscript{*}}

\usepackage{amsopn}

\ifpdf
\hypersetup{
  pdftitle={Sequential Design for Simulators with Sharp Variations via Deep Gaussian Process Gradients},
  pdfauthor={Y. Yang, D. Ming, and S. Guillas}
}
\fi

\begin{document}

\maketitle
\begin{abstract}
Deep Gaussian Processes (DGPs) compose Gaussian Process (GP) layers to warp inputs, improving the emulation of computer simulators with nonstationary input-output mappings. While gradient posteriors are analytically available for GPs, quantifying gradient uncertainty for DGPs is more challenging
 due to their hierarchical composition. In this paper, we propose an efficient approximation to the gradient distribution of a two-layer DGP emulator
without requiring simulator gradients. Using the local linearization, we derive closed-form approximations for the gradient mean and
covariance, enabling efficient gradient evaluation with uncertainty quantification (UQ). Empirical results show that the proposed approximation provides accurate gradient estimates and informative UQ on nonstationary examples. We then use the derived gradient uncertainty to guide sequential design for computer models with sharp variations. Specifically, we represent sharp-variation regions as super-level sets of the gradient norm and introduce an entropy-based acquisition rule that selects new samples where the estimated
sharp-region is most uncertain. Experiments on synthetic benchmarks and a real-world application show that the resulting sequential design
improves emulation accuracy for functions with sharp variations relative to existing design methods.
\end{abstract}



\section{Introduction}
Computer simulators are widely used to investigate complex scientific and engineering systems by mapping input parameters to output responses. In many applications, scientifically important regimes are localized in small regions of the input space where the response changes rapidly. These sharp-variation regions 
can correspond to qualitative changes in system properties, such as bifurcations and phase transitions. For example, in aerodynamics, changes in operating conditions, such as the Mach number, can move the flow into the transonic regime, where shock formation and shock-induced separation lead to steep changes in related responses such as lift, drag, and pressure \citep{dupuis2018surrogate}. Another example can be found in fluid dynamics: in Rayleigh--B\'enard convection, changes in physical parameters such as the Rayleigh and Prandtl numbers can induce transitions among regular convective and chaotic flow regimes, leading to sharp changes in convective heat-transfer efficiency \citep{lorenz1963deterministic,dullin2007extended,ahlers2009heat}. Because computer simulators are often computationally expensive, intensive exploration of the input space is infeasible. Statistical emulators provide a practical way to approximate response surfaces from a limited number of simulator runs. Constructing an accurate emulator in this setting requires both a reliable global approximation and attention to localized sharp-variation regions.

In computer experiments, space-filling designs are commonly used to provide broad coverage of the input domain \citep{santner2003design}. When the simulation budget is limited, sharp-variation regions may remain undetected under space-filling designs, as they often occupy only a small portion of the input domain. Sequential design can address this challenge by selecting new designs adaptively based on the current emulator. Classical variance-based criteria, including active learning MacKay (ALM) \citep{mackay1992information}, active learning Cohn (ALC) \citep{cohn1996active}, and mutual information \citep{krause2008near,beck2016sequential}, rely on Gaussian process (GP) emulators for new design selection. However, standard GPs with stationary kernels can struggle to represent abrupt variations, potentially reducing the efficiency of sequential design near sharp transitions. This motivates the development of nonstationary GPs. Existing nonstationary GP approaches can be grouped into three categories: nonstationary kernel designs; partition and local GPs; and deep Gaussian processes (DGPs). Nonstationary kernel methods increase flexibility by allowing the covariance structure to vary over the input space, often by introducing spatially varying hyperparameters or data-driven feature representations \citep{paciorek2006spatial,binois2018practical,mohammadi2019emulating,higdon2022non}. Partition and local GP methods, such as treed GPs \citep{gramacy2008bayesian}, local approximate GPs \citep{gramacy2015local}, and jump GPs \citep{park2022jump}, introduce nonstationary flexibility by applying GPs locally over learned partitions. In recent years, DGPs have become increasingly popular for nonstationary emulation. By stacking multiple GPs in a feed-forward composition, DGPs learn latent mappings that warp the input space and provide a flexible global modeling framework for representing nonstationary functions with varying smoothness and uncertainty \citep{damianou2013deep,dunlop2018deep,ming2023deep,sauer2023active}. 
\begin{figure}
    \vspace{-1em}
    \centering
    \includegraphics[width=0.9\linewidth]{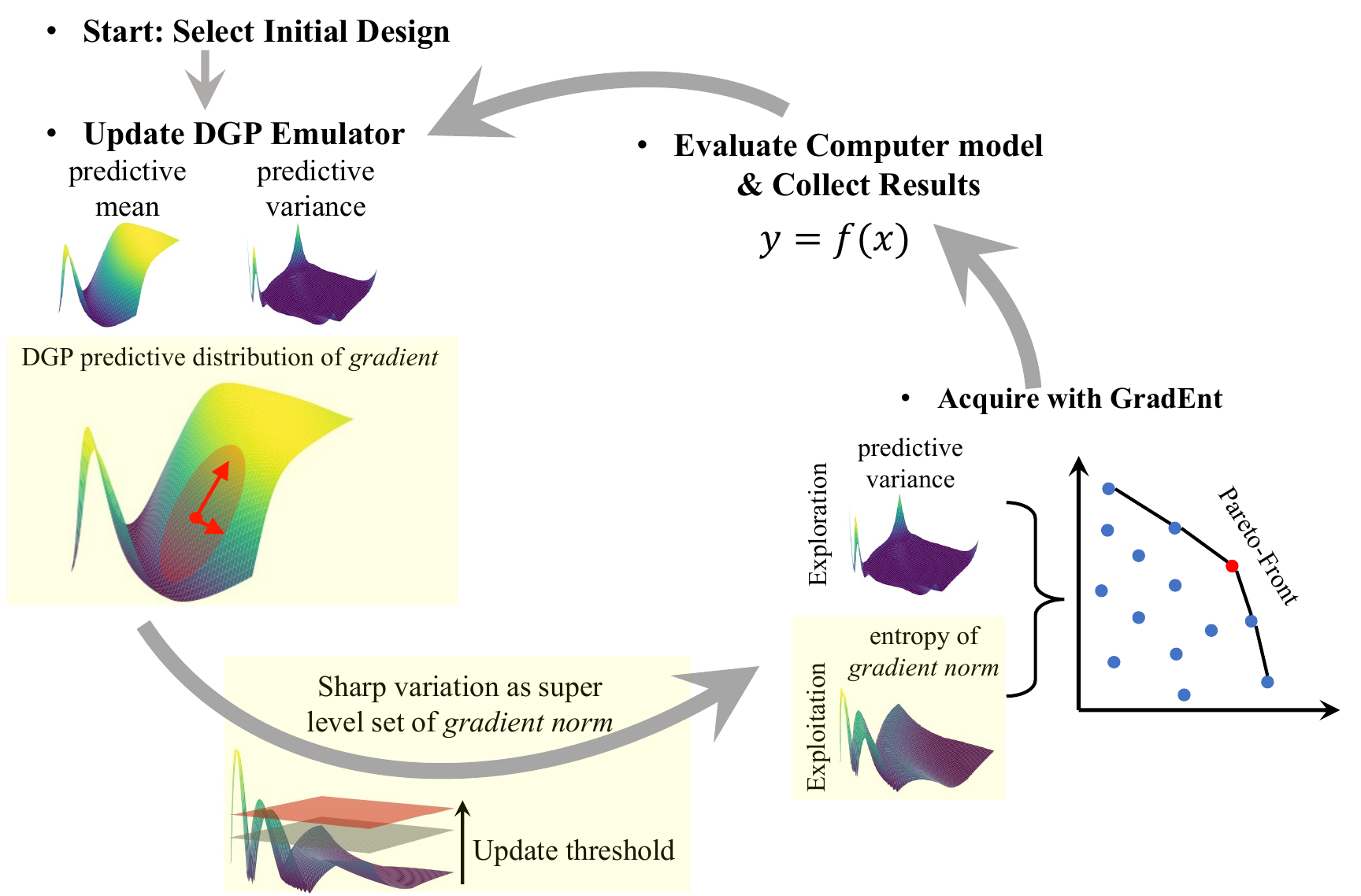}
    \vspace{-0.5em}
    \caption{An overview of our proposed methodology, with the main innovations shown in yellow. We first derive an efficient approximation to the predictive distribution of the DGP gradient. This gradient information is then incorporated in the proposed GradEnt acquisition function to guide an iterative sequential design process, efficiently focusing sampling on regions of sharp variation.}
    \label{fig:flow_chart_main}
    \vspace{-1.2em}
\end{figure}

Prior studies have advanced sequential design using nonstationary GPs. Early work combined treed GPs, which partition the input space into
axis-aligned regions, with the variance-based ALM and ALC criteria
\citep{gramacy2008bayesian}. More recently, ALC was incorporated with fully
Bayesian DGPs \citep{sauer2023active}. These approaches have shown promise for sequential emulation in various
applications, including aerodynamics \citep{gramacy2009adaptive}. Despite their different representations of nonstationarity, these
design criteria remain variance-based and do not explicitly account
for sharp variations. A region may have high
predictive variance because the existing design points are
locally sparse, even when the underlying response is smooth. Therefore, driven by general predictive uncertainty, such criteria may
not efficiently target localized sharp-variation regions. Jump GPs extend the partition-based framework of treed GPs for more flexible boundaries \citep{park2022jump}, and their
sequential design approach further combines predictive variance with
estimated model bias arising from uncertainty in partitions
\citep{park2025active}. This strategy is well suited to piecewise response surfaces with
clearly separated regimes. For continuous sharp variations,  partitioning may prevent
nearby observations across the transition from jointly inferring local predictions and thereby reduce sample efficiency. The bias formulation also assumes at most two response regions locally, which may limit extensions to complex multi-region junctions. These observations motivate a design criterion that targets sharp-variation regions without assuming a specific partition structure. We therefore develop GradEnt, a sequential design framework that utilizes
the DGP emulator to represent sharp variations without explicit
partitioning and its predictive gradient information to prioritize sampling in these regions.

The effectiveness of gradient-based design criteria induced by GP emulators has been explored in \citep{marmin2018warped}. In this work, since we use a DGP emulator, a key step is to derive the posterior distribution of DGP gradients.
We therefore build on the DGP framework in \citep{ming2023deep} and derive analytical approximations to the posterior gradient distributions of a two-layer DGP emulator. Moreover, we develop a sequential design algorithm that builds on the derived gradient distribution to form a probabilistic estimate of sharp-variation regions and guide sampling while supporting accurate global emulation. 

Figure \ref{fig:flow_chart_main} illustrates the high-level structure of our proposed methodology, where a DGP emulator and the proposed acquisition function work together in an iterative loop (the yellow sections indicate the methodological innovations of this paper). 

The paper is organized as follows. Section \ref{sec:background} reviews GP, LGP, and DGP emulators, providing the basis for the derivation of DGP gradients. Section \ref{sec: DGP gradient estimation} develops an efficient approximation to the DGP gradient posterior and validates its effectiveness numerically. This gradient approximation is then incorporated into the sequential design framework introduced in Section \ref{sec: Sequential Design}. Finally, Section \ref{Sec:experiment} evaluates the performance of the proposed sequential design method through synthetic experiments and a real-world application.

\section{Background}
\label{sec:background}
\paragraph{Notation} In this paper, we consider a computer model $f$ that maps a $d$ dimensional input $\boldsymbol{x}=[x_1,...,x_d]^\top\in\mathcal{X}\subset\mathbb{R}^d$ to a scalar output $y(\boldsymbol{x})\in\mathbb{R}$. Bold uppercase symbols $\boldsymbol{X}$ denote collections of inputs, such as $\boldsymbol{X}=[\boldsymbol{x}_1,...,\boldsymbol{x}_{n}]^\top\in\mathbb{R}^{n\times d}$, with corresponding outputs $\boldsymbol{y}(\boldsymbol{X})=[f(\boldsymbol{x}_1),...,f(\boldsymbol{x}_n)]^\top\in\mathbb{R}^n$ or simply $\boldsymbol{y}$. The bold characters generally indicate vectors or matrices formed by concatenating the relevant elements. $\|\cdot\|$ denotes the Euclidean norm. The notation is summarized in Appendix \ref{sec: notations}.

\subsection{Gaussian Process}
\label{Subsec: Gaussian Process}
Given $n$ $d$-dimensional inputs $\boldsymbol{X}\in\mathbb{R}^{n\times d}$, the GP model assumes that the output follows a multivariate Gaussian distribution, $\boldsymbol{y}(\boldsymbol{X})\sim\mathcal{N}\big(\boldsymbol{\mu}(\boldsymbol{X}),\boldsymbol{\Sigma}(\boldsymbol{X})\big)$ \cite{rasmussen2006gaussian}. We assume a zero-mean prior and a product-form kernel (e.g., squared exponential (SE) or Mat\'ern), which defines the correlation matrix $\boldsymbol{R}(\boldsymbol{X})$, where $\boldsymbol{R}(\boldsymbol{X})_{ij}=\prod_dk_d(\boldsymbol{x}_{id},\boldsymbol{x}_{jd})+\eta\mathbbm{1}_{\{i=j\}}$ and $\eta$ are the nugget terms. The covariance matrix is $\boldsymbol{\Sigma}(\boldsymbol{X})=c^2\boldsymbol{R}(\boldsymbol{X})$. These kernels include unknown length-scale parameters $\gamma_d$ that are estimated alongside $c^2$ and $\eta$. Given the GP parameters $\boldsymbol{\theta}=\{c^2, \eta, \boldsymbol{\gamma}\}$, the GP emulator is defined as the posterior distribution of $y(\boldsymbol{x}_*)$ at a new input $\boldsymbol{x}_*\in\mathbb{R}^d$ which is Gaussian with mean
and variance given by:
\begin{equation}
    \mu_{\text{GP}}(\boldsymbol{x}_*)=\boldsymbol{r}(\boldsymbol{x}_*)^\top\boldsymbol{R}(\boldsymbol{X})^{-1}\boldsymbol{y},\quad\sigma_{\text{GP}}^2(\boldsymbol{x}_*)=c^2\big(1+\eta-\boldsymbol{r}(\boldsymbol{x}_*)^\top\boldsymbol{R}(\boldsymbol{X})^{-1}\boldsymbol{r}(\boldsymbol{x}_*)\big),
    \label{eq:gp_posterior_predicitive}
\end{equation}
where $\boldsymbol{r}(\boldsymbol{x}_*)=[k(\boldsymbol{x}_{1},\boldsymbol{x}_*),...,k(\boldsymbol{x}_{n},\boldsymbol{x}_*)]^\top\in\mathbb{R}^{n}$.

\subsection{Linked Gaussian Process}
\label{subsec: linked gp}
Linked Gaussian process (LGP) models feedforward networks of computer models by linking GP emulators of individual models \cite{kyzyurova2018coupling,ming2021linked}.
Consider two computer models: the first model 
$\boldsymbol{f}\coloneqq(f_1,...f_p)$ maps a $d$-dimensional input $\boldsymbol{x}\in\mathbb{R}^{d}$ to a $p$-dimensional output $\boldsymbol{w}\in\mathbb{R}^{p}$, which then serves as input for the second model $g$,  producing a scalar output $y\in\mathbb{R}$.
\begin{wrapfigure}{r}
{0.5\textwidth}
\vspace{-0.5em}
\centering
\begin{tikzpicture}[scale=0.9, shorten >=1pt,->,draw=black!50, node distance=3cm]
    \tikzstyle{every pin edge}=[<-,shorten <=1pt]
    \tikzstyle{neuron}=[circle,fill=black!25,minimum size=30.5pt,inner sep=0pt]
    \tikzstyle{layer1}=[neuron, fill=green!50]
    \tikzstyle{layer2}=[neuron, fill=red!50]
    \tikzstyle{layer3}=[neuron, fill=blue!50]
    \tikzstyle{annot} = [text width=4em, text centered]
    \node[layer1, pin=left:$\boldsymbol{x}$] (I-0) at (0,0) {$\hat{f}_1$};
    \node[layer1, pin=left:$\boldsymbol{x}$] (I-1) at (0,-1.25) {$\hat{f}_2$};
    \node[layer1, pin=left:$\boldsymbol{x}$] (I-2) at (0,-2.95) {$\hat{f}_p$};
            \node[layer2, pin={[pin edge={->}]right:$y$}] (H-1) at (4cm,-1.25 cm) {$\hat{g}$};
            \path [draw] (I-0) -- (H-1) node[font=\large,pos=0.35,fill=white,align=left,sloped] {$w_{1}$};
            \path [draw] (I-1) -- (H-1) node[font=\large,pos=0.35,fill=white,align=left,sloped] {$w_{2}$};
            \path [draw] (I-2) -- (H-1) node[font=\large,pos=0.35,fill=white,align=left,sloped] {$w_p$};
            \path (I-1) -- (I-2) node [black, midway, sloped] {$\dots$};
            \path (I-1) -- (I-2) node [black, midway, sloped,transform canvas={xshift=-12.5mm}] {$\dots$};
            \path (I-1) -- (I-2) node [black, midway, sloped,transform canvas={xshift=15mm,yshift=3mm}] {$\dots$};
\end{tikzpicture}
\vspace{-2em}
\caption{The hierarchy of GPs forms an LGP emulator representing a feed-forward system of two computer models.}
\label{fig:linked_gp_structure}
\vspace{-1em}
\end{wrapfigure}
Let $\hat{\boldsymbol{f}}$ and $\hat{g}$ denote the GP emulators for the first and second models, respectively. Assuming $\boldsymbol{w}$ is conditionally independent across its $p$ dimensions given $\boldsymbol{x}$, the first emulator $\hat{\boldsymbol{f}}$ consists of $p$ independent GPs, $\{\hat{f}_i\}_{i=1}^p$, where each $\hat{f}_i$ is a GP mapping $\boldsymbol{x}$ to the corresponding output $w_i$. The LGP emulator then connects these individual GP emulators to form a hierarchical emulator, as illustrated in Figure~\ref{fig:linked_gp_structure}.

Under the assumption of conditional independence, the intractable posterior predictive distribution of the global output can be sufficiently approximated by a Gaussian distribution with \textit{analytical} and \textit{exact} mean and variance of the predictive posterior $y(\boldsymbol{x}_*)$. This Gaussian approximation defines the LGP emulator, with the predictive mean and variance given by
\begin{equation}
    \begin{split}
        \mu_{\text{LGP}}(\boldsymbol{x}_*)=&\boldsymbol{I}(\boldsymbol{x}_*)^\top\boldsymbol{R}(\boldsymbol{W})^{-1}\boldsymbol{y} \\
        \sigma_{\text{LGP}}^2(\boldsymbol{x}_*)=&\boldsymbol{y}^\top\boldsymbol{R}(\boldsymbol{W})^{-1}\boldsymbol{J}(\boldsymbol{x}_*)\boldsymbol{R}(\boldsymbol{W})^{-1}\boldsymbol{y}-(\boldsymbol{I}(\boldsymbol{x}_*)^\top\boldsymbol{R}(\boldsymbol{W})^{-1}\boldsymbol{y})^2 \\
        &+\sigma^2(1+\eta-\text{tr}[\boldsymbol{R}(\boldsymbol{W})^{-1}\boldsymbol{J}(\boldsymbol{x}_*)])
    \end{split}
\label{eq: LGP analytical mean and variance}
\end{equation}
where (1) $\boldsymbol{I}(\boldsymbol{x}_*)\in\mathbb{R}^{n}$ has entries $I_i(\boldsymbol{x}_*)=\mathbb{E}[k(\boldsymbol{w}(\boldsymbol{x}_*),\boldsymbol{w}_i)]$; $\boldsymbol{w}(\boldsymbol{x}_*)$ are the outputs of the first-layer emulators $\hat{\boldsymbol{f}}$ and $\boldsymbol{w}_i$ is the $i$th observed internal sample. For brevity, we write \(\boldsymbol{w}_*=\boldsymbol{w}(\boldsymbol{x}_*)\). Equivalently,  
\(
I_i(\boldsymbol{x}_*)=\int k(\boldsymbol{w},\boldsymbol{w}_i)\,p(\boldsymbol{w}\mid\boldsymbol{x}_*)\,d\boldsymbol{w}.
\)  
Thus, $\boldsymbol{I}(\boldsymbol{x}_*)$ is a convolution of $k$ with the first-layer predictive distribution, acting as a nonstationary kernel in the LGP framework. (2) $\boldsymbol{J}(\boldsymbol{x}_*)\in\mathbb{R}^{n\times n}$ with its $(i,j)$ entry $J_{ij}(\boldsymbol{x}_*)=\prod_{p}\mathbb{E}\big[k_p\big(w_p(\boldsymbol{x}_*),w_{i,p}\big)k_p\big(w_p(\boldsymbol{x}_*),w_{j,p}\big)\big]$. The $\boldsymbol{I}(\boldsymbol{x}_*)$ and $\boldsymbol{J}(\boldsymbol{x}_*)$ can be expressed in closed form for a broad class of kernel functions \cite{kyzyurova2018coupling,ming2021linked}. This analytical Gaussian approximation has been shown to be effective in terms of minimizing the Kullback–Leibler (KL) divergence \cite{ming2023deep}.

\subsection{Deep Gaussian Process}
\label{Subsec: DGP}
Deep Gaussian Process (DGP) emulators can be viewed as an LGP with \textit{unobservable} internal Input/Output (I/O) $\boldsymbol{W}$. For example, Figure \ref{fig:linked_gp_structure} represents a two-layered DGP when the training data $\boldsymbol{W}$ is unobservable. Such unobservable variables pose challenges for efficient training and inference. Many solutions have been proposed, including MCMC-based methods \cite{sauer2023active} and variational-inference methods \cite{bui2016deep,dunlop2018deep}. In this paper, we instead adopt Stochastic Imputation (SI) \cite{ming2023deep}, a Stochastic EM–based method \cite{celeux1985sem} that exploits structural similarities between DGPs and LGPs. Given the training data $\{\boldsymbol{X},\boldsymbol{y}\}$, the SI method iterates between (i) \textit{imputation}, drawing $\boldsymbol{W}$ from
$p(\boldsymbol{W}\mid\boldsymbol{X},\boldsymbol{y})\propto
p(\boldsymbol{y}\mid\boldsymbol{W},\boldsymbol{\theta}_g)
p(\boldsymbol{W}\mid\boldsymbol{X},
\{\boldsymbol{\theta}_i\}_{i=1}^p)$, and (ii) \textit{maximization}, updating
$\boldsymbol{\theta}_g$ and $\{\boldsymbol{\theta}_i\}_{i=1}^p$
conditional on $\boldsymbol{W}$ by maximizing the corresponding GP likelihoods.

To account for the imputation uncertainties, one can draw $N_{\text{imp}}$ realizations from $p(\boldsymbol{W}|\boldsymbol{y},\boldsymbol{X},\{\boldsymbol{\theta}_i\}_{i=1}^p)$ and construct $N_{\text{imp}}$ LGPs. This yields $N_{\text{imp}}$ MC approximations of the DGP, whose posterior predictive mean and variance are given by 
\begin{equation}
\vspace{-0.5em}
\begin{split}
    &\mu_{\text{DGP}}(\boldsymbol{x}_*)=\frac{1}{N_{\text{imp}}}\sum_{i=1}^{N_{\text{imp}}}\mu_{\text{LGP},i}(\boldsymbol{x}_*),\\
    &\sigma_{\text{DGP}}^2(\boldsymbol{x}_*)=\frac{1}{N_{\text{imp}}}\sum_{i=1}^{N_{\text{imp}}}\text{\large(}\mu_{\text{LGP},i}(\boldsymbol{x}_*)^2+\sigma_{\text{LGP},i}^2(\boldsymbol{x}_*)\text{\large)}-\mu_{\text{DGP}}(\boldsymbol{x}_*)^2,
\end{split}
    \vspace{-0.5em}
    \label{eq: DGP predicitive distribution}
\end{equation}
where $\{\mu_{\text{LGP},i},\sigma^2_{\text{LGP},i}\}_{i=1}^{N_{\text{imp}}}$ are the mean and variance of the $i$th LGP (see Equation~\eqref{eq: LGP analytical mean and variance}). Further algorithmic details for SEM training and prediction are provided in supplementary material (SM) Section 1.

\subsection{Gradient of Gaussian Process Emulator}
\label{subsec: gradient_of_gaussian_process_emulator}
Another key property of the GP is that any linear operation of a GP remains a GP \cite{rasmussen2006gaussian}. Given a GP emulator $\hat{f}$, the gradient $\nabla\hat{f}(\boldsymbol{x}_*)\in\mathbb{R}^d$ has the $i$th component \(\partial_{x_{*,i}} \hat{f}(\boldsymbol{x}_*)
\). The posterior distribution of $\nabla\hat{f}$ remains a Gaussian distribution with mean $\mu_\nabla(\boldsymbol{x}_*)\in\mathbb{R}^d$ and covariance $\Sigma_{\nabla}(\boldsymbol{x}_*)\in\mathbb{R}^{d\times d}$, given as \cite{solak2002derivative,mchutchon2015nonlinear,rasmussen2006gaussian}:
\begin{equation}       \mu_\nabla(\boldsymbol{x}_*)=\nabla\boldsymbol{r}(\boldsymbol{x}_*)^\top\boldsymbol{R}(\boldsymbol{X})^{-1}\boldsymbol{y},\quad\Sigma_\nabla(\boldsymbol{x}_*)=c^2\big(H-\nabla\boldsymbol{r}(\boldsymbol{x}_*)^\top\boldsymbol{R}(\boldsymbol{X})^{-1}\nabla\boldsymbol{r}(\boldsymbol{x}_*)\big),
\label{eq:gradient_of_gp}
\end{equation}
where $\nabla\boldsymbol{r}(\boldsymbol{x}_*) \in \mathbb{R}^{n\times d}$ is the Jacobian matrix of the kernel vector, with the $(i,j)$-th entry given by $(\nabla\boldsymbol{r})_{ij} = \frac{\partial k(\boldsymbol{x}_i, \boldsymbol{x}_*)}{\partial x_{*,j}}$. Here, $H=\left.
\nabla_{\mathbf{u}}\nabla_{\mathbf{v}}^\top
k(\mathbf{u},\mathbf{v})
\right|_{\mathbf{u}=\mathbf{v}=\boldsymbol{x}_*}
\in\mathbb{R}^{d\times d}$\footnote{The kernel $k(\cdot,\cdot)$ is at least twice differentiable for $H$ to exist, a property satisfied by common kernels such as the squared exponential and Matérn-2.5.} denotes the matrix of kernel cross-derivatives evaluated at
$\boldsymbol{x}_*$.

\section{Posterior Distribution of DGP Emulator Gradients}
\label{sec: DGP gradient estimation}
As shown in Section~\ref{subsec: gradient_of_gaussian_process_emulator}, the gradients of GP emulators are straightforward to obtain, whereas extending this construction to nonstationary DGP emulators is challenging. In this section, we develop an efficient approximation to the posterior distribution of DGP gradients. 

A related line of work studies gradient-enhanced DGP emulators, where simulator gradients are observed and incorporated as additional training data \citep{booth2025deep}. Differently, we consider the common black-box computer experiment setting, where simulators return only function values. This motivates an emulator-based construction of gradient uncertainty. Our method builds on the SI scheme, which exploits the structural similarity between LGPs and DGPs to enable a tractable approximation. In Section~\ref{subsec: DGP gradient estimation}, we first derive a closed-form Gaussian approximation to the LGP gradient posterior with an exact predictive expectation. Under SI, the DGP emulator is represented as an ensemble of imputed LGPs. We apply the proposed approximation to each LGP realization and aggregate the resulting gradient posteriors, yielding a Monte Carlo (MC) approximation of the DGP gradient posterior.

\subsection{Posterior of LGP gradient} 
\label{subsec: DGP gradient estimation}
Consider the two-layered LGP in the form of Figure \ref{fig:linked_gp_structure}; The global output is \(y(\boldsymbol{x}_*)=\hat{g}(\hat{\boldsymbol{f}}(\boldsymbol{x}_*))\), and its gradient follows from the multivariate chain rule:
\begin{equation}
    \label{eq: LGP gradient chain rule}\nabla_{\boldsymbol{x}_*}y(\boldsymbol{x}_*)=\nabla_{\boldsymbol{w}_*}\hat{g}(\boldsymbol{w}_*)^\top\text{Jac}_{\boldsymbol{x}_*}\hat{\boldsymbol{f}}(\boldsymbol{x}_*),
\end{equation}
where $\text{Jac}_{\boldsymbol{x}_*}\hat{\boldsymbol{f}}(\boldsymbol{x}_*)=[\nabla_{\boldsymbol{x}_*}\hat{f}_1(\boldsymbol{x}_*),...,\nabla_{\boldsymbol{x}_*}\hat{f}_p(\boldsymbol{x}_*)]^\top\in\mathbb{R}^{p\times d}$ is the Jacobian of the first layer GP emulators with respect to the inputs $\boldsymbol{x}_*$. Because both layers are stochastic and nonlinearly coupled through $\boldsymbol{w}_*$, the exact posterior of $\nabla_{\boldsymbol{x}_*} y(\boldsymbol{x}_*)$ is analytically intractable. We therefore first derive the posterior mean of the gradient under the LGP construction and then use this as the mean of a Gaussian approximation to the exact gradient distribution.

The next lemma shows that, for a two-layer LGP, the posterior mean of the global gradient equals the gradient of the LGP posterior mean.
\begin{lemma}
    \label{lemma: exactness by differentiating mean}
    Consider the two-layer LGP as in Section~\ref{subsec: linked gp}:
    \begin{itemize}
        \item First layer: $p$ independent GPs \(\hat{f}_i\) with squared exponential kernels, trained on $(\boldsymbol{X},\boldsymbol{W})$. At a new input $\boldsymbol{x}_*$, the predictive distribution is 
        \(w_i(\boldsymbol{x}_*)|\mathcal{D}\sim\mathcal{N}(\mu_i(\boldsymbol{x}_*), \sigma_i^2(\boldsymbol{x}_*)).\) for $i=1,...,p$.
        \item Second layer: a GP $\hat{g}$ with a squared exponential kernel, trained on $(\boldsymbol{W},\boldsymbol{y})$.
    \end{itemize}
    Let $y(\boldsymbol{x}_*)=\hat{g}(\hat{\boldsymbol{f}}(\boldsymbol{x}_*))$ be the global random output, and let $\mu_{\text{LGP}}(\boldsymbol{x}_*)$ be the LGP predictive mean from \Cref{eq: LGP analytical mean and variance}. Then, we have, 
    \begin{equation}
        \label{eq:mean_derivative_of_lgp}
        \mathbb{E}[\nabla_{\boldsymbol{x}_*} y(\boldsymbol{x}_*)]=\nabla_{\boldsymbol{x}_*}\mu_{\text{LGP}}(\boldsymbol{x}_*)=\nabla\boldsymbol{I}(\boldsymbol{x}_*)^\top\boldsymbol{R}(\boldsymbol{W})^{-1}\boldsymbol{y},\quad \text{for every } \boldsymbol{x}_*\in\mathcal{X}.
    \end{equation}
\end{lemma}
In other words, under the LGP construction, the posterior mean of the global gradient is obtained exactly by differentiating the LGP predictive mean. The lemma therefore provides an efficient expression for this posterior mean. The proof relies on (i) conditional independence between layers and (ii) a reparameterization of the first-layer predictive distribution, which together justify exchanging the gradient and expectation. The full proof and the closed-form expression for $\nabla_{\boldsymbol{x}_*}\mu_{\mathrm{LGP}}(\boldsymbol{x}_*)$ under SE kernels are given in Appendix~\ref{subsec: Proof of Lemma 3.1}. The matrix $\nabla\boldsymbol{I}(\boldsymbol{x}_*)\in\mathbb{R}^{n\times d}$, with entries $[\nabla\boldsymbol{I}(\boldsymbol{x}_*)]_{ij}=\frac{\partial \boldsymbol{I}_i(\boldsymbol{x}_*)}{\partial x_{*,j}}$, quantifies the sensitivity of the $w_i$ to the $j$th input dimension. Since $\boldsymbol{I}(\boldsymbol{x}_*)$ is the expectation over the intermediate outputs $\boldsymbol{w}(\boldsymbol{x}_*)$, differentiation with respect to $\boldsymbol{x}_*$ implicitly applies the multivariate chain rule and propagates sensitivities from inputs through the hierarchy.

\paragraph{Approximation of the Covariance} 
The nonlinear product in \Cref{eq: LGP gradient chain rule} makes the exact covariance of the LGP gradient analytically intractable: the second-layer GP gradient is evaluated at the latent random variable \(\boldsymbol{w}_*\) and multiplied by the random Jacobian of the first-layer GPs. A straightforward approach is to estimate LGP gradients by aggregating numerical gradients of posterior samples. This would require MC sampling through the latent variable \(\boldsymbol{w}_*\) at each input point, making repeated evaluations over large input sets expensive, as in sequential design. To handle this nonlinear propagation efficiently, we apply the \textit{delta method} \cite{oehlert1992note}. Specifically, we linearize $\nabla_{\boldsymbol w}\hat g(\boldsymbol w)$ around the mean $\mu_{\boldsymbol w}$ via a first-order Taylor expansion $\nabla_{\boldsymbol{w}_*} \hat{g}(\boldsymbol{w}_*)= \nabla_{\mu_{\boldsymbol{w}_*}} \hat{g}(\mu_{\boldsymbol{w}_*})+O(\|\boldsymbol{w}_*-\mu_{\boldsymbol{w}_*}\|)$. The residual term represents higher-order uncertainties. By Jensen’s inequality, the expected magnitude satisfies
\(
\mathbb{E}\left[\|\boldsymbol{w}_* - \mu_{\boldsymbol{w}_*}\|\right] \le \sqrt{\mathbb{E}\left[\|\boldsymbol{w}_* - \mu_{\boldsymbol{w}_*}\|^2\right]} = \sqrt{\sum_{i=1}^p \sigma_i^2(\boldsymbol{x}_*)},
\)
where $\sigma_i^2(\boldsymbol{x}_*)$ is the posterior variance of the $i$th first-layer GP. If the kernel has smoothness $s$, then on bounded domains
\(
\sigma_i^2(\boldsymbol{x}_*) = \mathcal{O}(n^{-2s/d}) \Rightarrow \mathbb{E}\left[\|\boldsymbol{w}_* - \mu_{\boldsymbol{w}_*}\|\right] \le \mathcal{O}(n^{-s/d})
\)
with $n$ training points in dimension $d$. Thus, the delta-method error decays polynomially in $n$, with faster rates for smoother kernels. For further details, we refer readers to Theorem 5.4 and Corollary 5.5 in \cite{kanagawa2018gaussian}, and Section 11 of \cite{wendland2004scattered}.

For clarity, we omit the dependence on $\boldsymbol{x}_*$ and denote $\mu_\nabla^{\hat{g}}$ and $\Sigma^{\hat{g}}_\nabla$ as the predictive mean and covariance of the second-layer GP's gradient at $\mu_{\boldsymbol{w}_*}$, while $\mu_\nabla^{\hat{f}_i}$ and $\Sigma^{\hat{f}_i}_\nabla$ are the quantities for the $i_{\text{th}}$ GP in the first layer at $\boldsymbol{x}_*$. This linearization allows for an analytical approximation of the gradient covariance. The final expression decomposes the total gradient uncertainty into two key sources of uncertainty. The first summation represents the uncertainty from the first layer ($\Sigma^{\hat{f}_i}_\nabla$), which is scaled by the expected squared magnitude of the second-layer gradient (the sum of its squared mean and variance). This captures how the second-layer GP transforms the uncertainty of the first layer. The second summation represents the first layer's mean structure ($\mu_{\nabla}^{\hat{f}}\mu_{\nabla}^{\hat{f}\top}$) scaled by the gradient covariance of the second layer ($\Sigma^{\hat{g}}_{\nabla}$). This captures how the second layer's uncertainty is propagated along the directions of the first layer's mean gradients. The resulting approximation is:
\begin{equation}
    \label{eq: LGP gradient covariance}
    \Sigma_{\nabla,\text{LGP}}\approx\sum_{i=1}^p\left[ \left[\left(\mu^{\hat{g}}_{\nabla}\right)_i^2+\left(\Sigma^{\hat{g}}_{\nabla}\right)_{ii}\right]\Sigma^{\hat{f}_i}_\nabla\right] + \sum_{k=1}^p \sum_{l=1}^p (\Sigma^{\hat{g}}_{\nabla})_{kl} \mu_{\nabla}^{\hat{f}_k} (\mu_{\nabla}^{\hat{f}_l})^T,
\end{equation}
Explicit expressions for these GP gradient statistics are given in Equation~\ref{eq:gradient_of_gp}. The explicit derivation of Equation~\ref{eq: LGP gradient covariance} is provided in Appendix~\ref{subsec: Derivation of Covariance approximation of LGP gradient - Equation 3.2}; see Equation~\ref{eq: full derivation of covaraince approximation}.

\subsection{Approximation of the DGP Gradient Posterior
} Since a DGP emulator is a collection of LGPs, the predictive posterior mean and covariance of the DGP gradients can be approximated by averaging the results from $N_{\text{imp}}$ imputed LGPs using Equations \eqref{eq:mean_derivative_of_lgp} and \eqref{eq: LGP gradient covariance} with the law of total variance:
\begin{equation}
\label{eq: DGP gradient covariance} 
\begin{aligned}
    \mu_{\nabla,\text{DGP}}
    \approx \frac{1}{N_{\text{imp}}}\sum_{i=1}^{N_{\text{imp}}}
        \mu^i_{\nabla,\text{LGP}}, \\[6pt]
    \Sigma_{\nabla,\text{DGP}}
    \approx\frac{1}{N_{\text{imp}}}\sum_{i=1}^{N_{\text{imp}}}
        \left[\Sigma^i_{\nabla,\text{LGP}}+\mu_{\nabla,\text{LGP}}^i{\mu_{\nabla,\text{LGP}}^i}^\top \right] 
        & - \mu_{\nabla,\text{DGP}}\mu_{\nabla,\text{DGP}}^\top.
\end{aligned}
\end{equation}
where the superscript $i$ denotes the gradient moments from the $i$th imputed LGP.
\subsection{Numerical validation} 
\begin{wrapfigure}{r}{0.4\textwidth}
    \setlength{\intextsep}{6pt}
    \setlength{\columnsep}{10pt}
    \centering
    \vspace{-0.5em}
    \includegraphics[width=\linewidth]{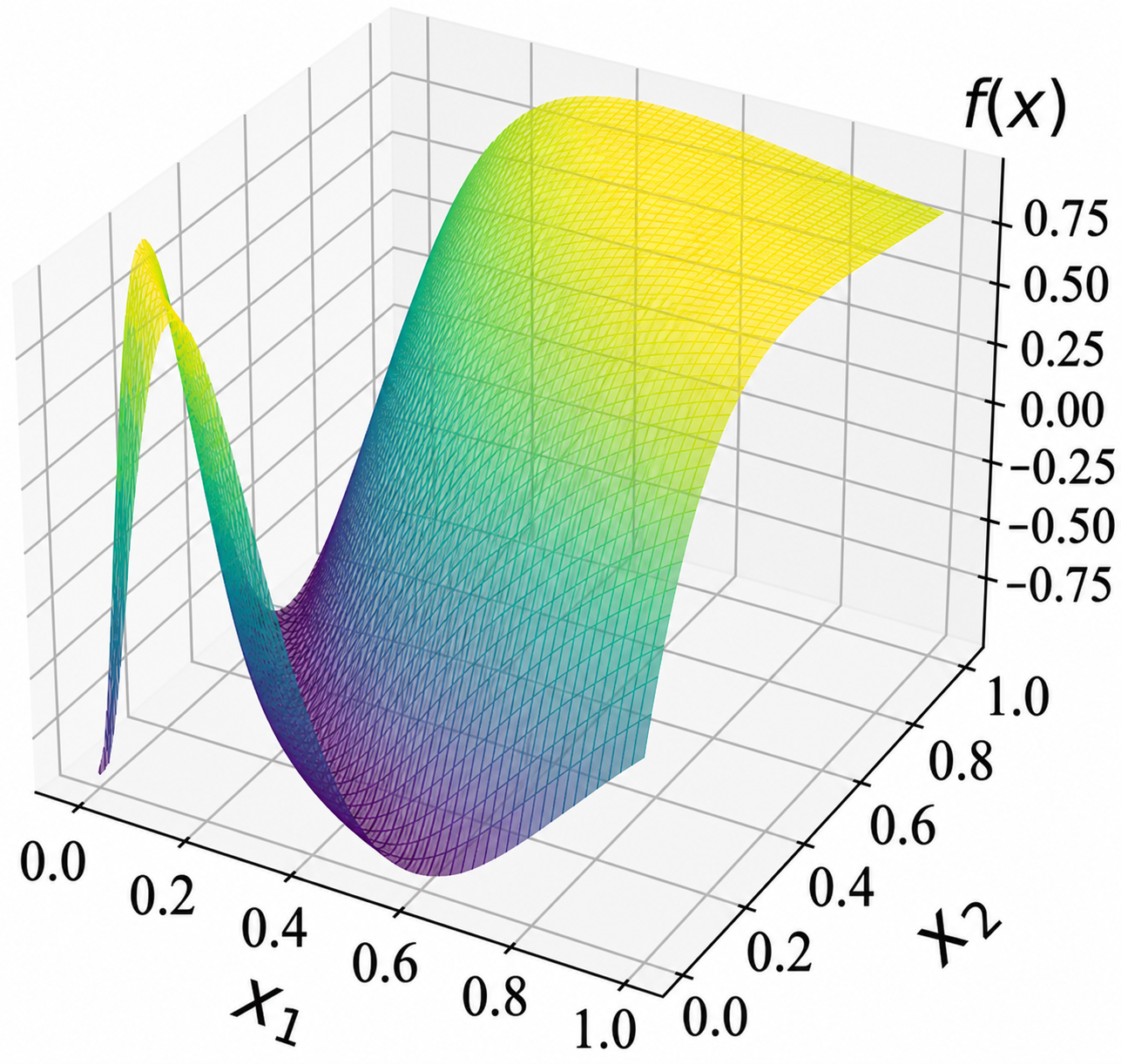}
    \vspace{-1.6em}
    \caption{3D visualization of 2D nonstationary wavy function}
    \vspace{-5em}
    \label{fig: 3d visualization of wavy function}
\end{wrapfigure}

This subsection validates the proposed DGP gradient approximation, referred to as DGP-Grad, using the following nonstationary wavy function with known gradients for $i=1,...,d$:
\begin{equation}
    \begin{split}
    f(\boldsymbol{x})=\sin{\Big(\frac{1}{\prod_{i=1}^d(0.7x_i+0.3)}\Big)},\quad \frac{\partial f}{\partial x_i}=-\frac{0.7\cos{\Big(\frac{1}{\prod_{i=1}^d(0.7x_i+0.3)}\Big)}}{(0.7x_i+0.3)\prod_{j=1}^d(0.7x_j+0.3)}.
\end{split}
\end{equation}
The wavy function is widely used as a nonstationary test case (see Section 5.1 in \cite{montagna2016computer} and Section 4.1 in \cite{volodina2020diagnostics}). Figure~\ref{fig: 3d visualization of wavy function} shows the \(d=2\) surface. We consider input dimensions \(d=1,2,3,4\), with training sets of sizes \(10,25,100,\) and \(300\), and test sets of sizes \(200,400,600,\) and \(800\), respectively, all generated by Latin hypercube design (LHD). 
We compare three gradient estimators: DGP-Grad, DGP-FD, and GP. DGP-Grad and DGP-FD use the same fitted two-layer DGP emulator with two first-layer GP nodes, each using an SE kernel. DGP-Grad computes gradients using the proposed posterior approximation. DGP-FD applies second-order central differences \citep{fornberg1988generation} with step size \(h=10^{-6}\) to the predictive mean of the same fitted DGP emulator, using a local \(3^d\) stencil around each evaluation point. The GP baseline uses an SE kernel and its analytical gradient posterior from Section~\ref{subsec: gradient_of_gaussian_process_emulator}.

We evaluate prediction accuracy using the mean relative error (MRE), defined as \(\text{MRE}=\frac{1}{n}\sum_{i=1}^n \|\nabla f(\boldsymbol{x}_i)-\mu_{\nabla}(\boldsymbol{x}_i)\|/\|\nabla f(\boldsymbol{x}_i)\|\). The gradient UQ is evaluated by the normalized dimension-wise continuous ranked probability score (NCRPS$_{\nabla}$),
\begin{equation}
\label{eq:ncrps-gradient}
\operatorname{NCRPS}_{\nabla}
=
\frac{1}{nd}\sum_{i=1}^n\sum_{j=1}^d
\frac{1}{\sigma_{\nabla,j}}
\int
\left(
F^\nabla_{i,j}(z)
-
\mathbb{I}\{z>[\nabla f(\boldsymbol{x}_i)]_j\}
\right)^2
\mathrm{d}z ,
\end{equation}
where $F^\nabla_{i,j}$ is the predictive cumulative distribution function (CDF) of the $j$th gradient component at $\boldsymbol{x}_i$. The scale parameter $\sigma_{\nabla,j}$ is the empirical standard deviation of the derivatives along dimension $j$. NCRPS$_\nabla$ assesses each marginal predictive distribution
against the corresponding true gradient component but does not capture cross-component dependence.
\begin{figure}[ht]
    \centering
    \includegraphics[width=0.9\linewidth]{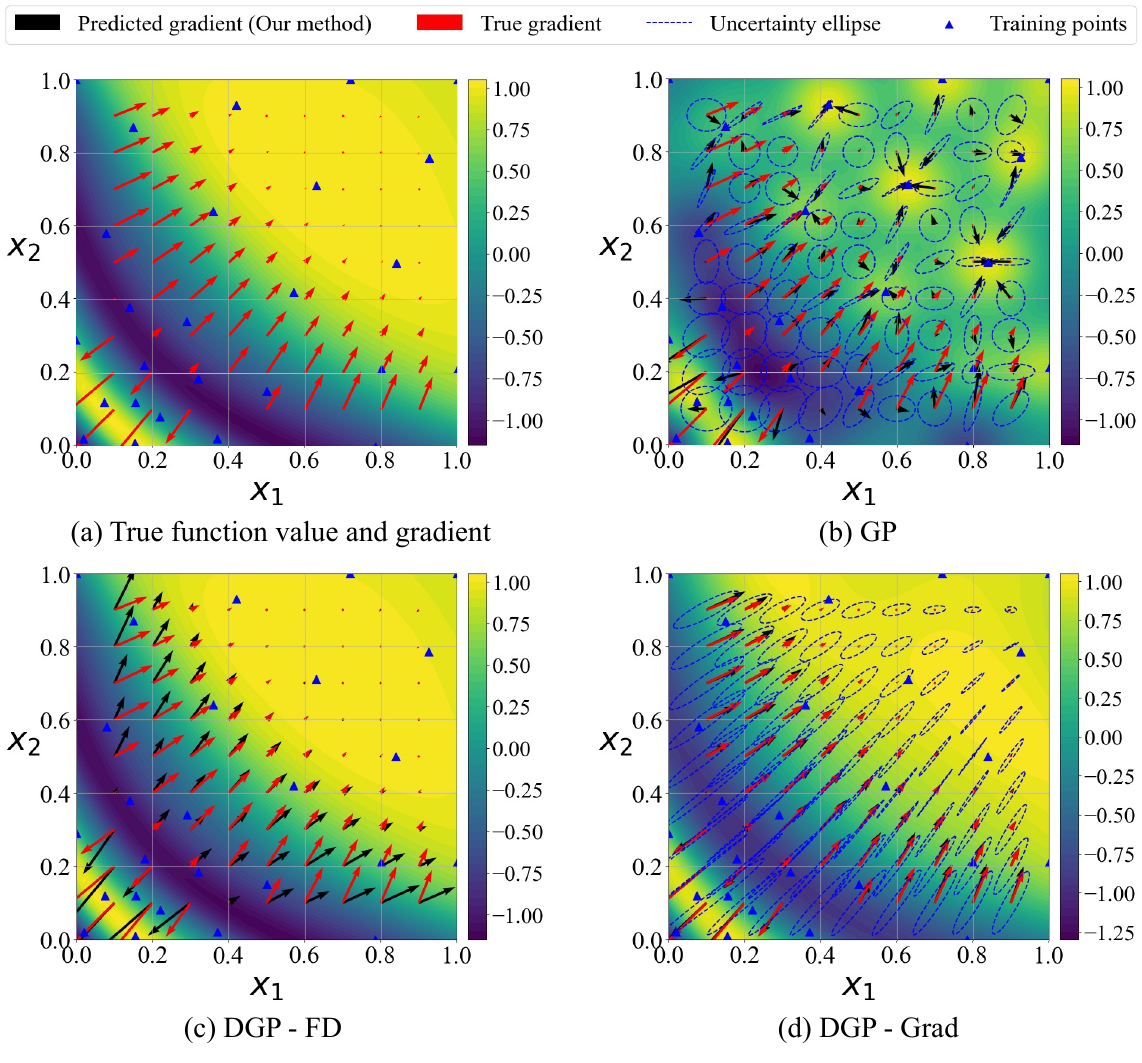}
    \vspace{-1em}
    \caption{(a) True function values (contours) and gradients. (b) Predicted values and gradients from a GP emulator. (c) FD-based gradient estimates on the predictive mean of the DGP emulator. (d) Posterior gradient distribution for the same DGP using the proposed method in \eqref{eq: DGP gradient covariance}.}
    \label{fig: comparsion_of_gradient}
    \vspace{-0.5em}
\end{figure}

Table \ref{tab: grad_comp} summarizes the quantitative results. DGP-Grad consistently achieves the lowest MRE across all dimensions, indicating a more accurate gradient mean prediction than both GP and DGP-FD. For UQ, GPs attain the lowest $\operatorname{NCRPS}_{\nabla}$ in \(d=1\) and \(d=4\), suggesting tighter gradient uncertainty in these cases, whereas DGP-Grad achieves the best NCRPS$_\nabla$ in \(d=2\) and \(d=3\). The higher NCRPS$_\nabla$ of DGP-Grad in $d=1,4$ could be attributed to conservative uncertainty propagation across latent layers, which can widen the predictive gradient distribution even when the gradient mean is accurate.  Overall, DGP-Grad improves gradient mean accuracy while providing informative UQ for gradient estimates.

Figure \ref{fig: comparsion_of_gradient} provides a qualitative comparison for \(d=2\). The GP emulator struggles to accurately emulate the function and, therefore, the gradient. While DGP-FD performs well in the flat upper-right region, it lacks robustness near regions of sharp response variation due to its sensitivity to the finite-difference step size \(h\) and discretization \citep{quarteroni2010numerical}. Figure \ref{fig: comparsion_of_gradient} (d) presents the results of our proposed method. The blue dashed ellipse represents the predictive uncertainty of the gradient, with larger variance along directions where the response changes more rapidly.

\begin{table}[ht]
\centering
\caption{Quantitative performance comparison of gradient computation methods across dimensions $d=1,2,3,4$. Results are reported as mean $\pm$ standard deviation over 40 replicates. For both MRE and NCRPS$_\nabla$, smaller values are better. "-" denotes metrics that are not applicable.}
\label{tab: grad_comp}
\footnotesize
\setlength{\tabcolsep}{6pt}
\renewcommand{\arraystretch}{1.0}
\begin{tabular}{lccc|ccc}
\toprule
& \multicolumn{3}{c|}{MRE} & \multicolumn{3}{c}{$\text{NCRPS}_{\nabla}$} \\
\cmidrule(lr){2-4}\cmidrule(lr){5-7}
$d$ & GP & DGP-FD & DGP-Grad & GP & DGP-FD & DGP-Grad \\
\midrule
1 & 0.25 $\pm$ 0.12 & 0.08 $\pm$ 0.03 & \textbf{0.07 $\pm$ 0.02}
  & \textbf{0.30 $\pm$ 0.05} & -- & 0.32 $\pm$ 0.09 \\
2 & 0.42 $\pm$ 0.15 & 0.16 $\pm$ 0.07 & \textbf{0.15 $\pm$ 0.06}
  & 0.32 $\pm$ 0.07 & -- & \textbf{0.24 $\pm$ 0.06} \\
3 & 0.55 $\pm$ 0.14 & 0.29 $\pm$ 0.12 & \textbf{0.26 $\pm$ 0.10}
  & 0.35 $\pm$ 0.10 & -- & \textbf{0.31 $\pm$ 0.11} \\
4 & 0.63 $\pm$ 0.10 & 0.45 $\pm$ 0.15 & \textbf{0.41 $\pm$ 0.14}
  & \textbf{0.40 $\pm$ 0.12} & -- & 0.45 $\pm$ 0.16 \\
\bottomrule
\end{tabular}
\end{table}

\section{Sequential Design - GradEnt}
\label{sec: Sequential Design}
In this section, we consider computer model outputs with sharp variations, where large gradient magnitudes indicate critical transitions. Such regions are challenging for stationary emulators and for sequential design criteria driven only by predictive uncertainty. To address this setting, we integrate DGP emulation with sequential design and propose \textit{GradEnt}, an entropy-based acquisition criterion that uses DGP gradient uncertainty to guide sampling near sharp-variation regions. Predictive uncertainty is incorporated as a second objective in a Pareto-frontier framework to support global emulation accuracy.

\subsection{Sharp Variation Region}
\label{subsec: Sharp Variation Region}
\begin{figure}
    \centering
    \includegraphics[width=1\textwidth]{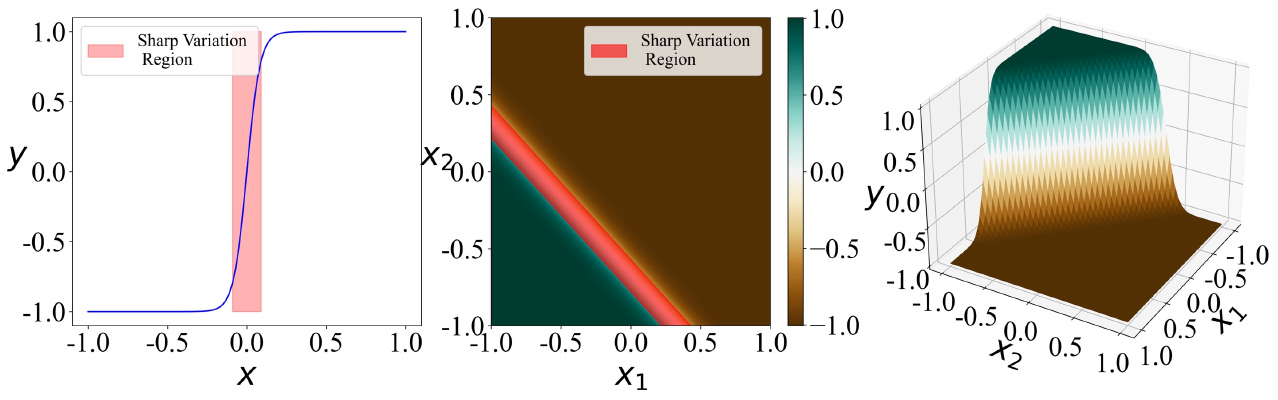}
    \vspace{-1.5em}
    \caption{Demonstration of high variation region indicated by the super level set of gradient norm $D_L=\{\boldsymbol{x}\in\mathcal{X}\colon\|\nabla f(\boldsymbol{x})\|\ge L\}$. The red shadow indicates the region $D$. (Left): The $1$-d plateau function with threshold $L=4$. (Middle): Contour plot of $2$-d plateau function with threshold $L=5$. (Right): $3$-d surface plot of the $2$-d plateau function}
    \label{fig:high_variation_region}
    \vspace{-0.5em}
\end{figure}

We first define the sharp-variation region used in the proposed sequential design. It provides an operational representation of locations where the response exhibits rapid change. Since the true function is unknown, this region is estimated from the emulator and updated sequentially as new design points are added. Following the weak-discontinuity definition in \cite{audet2022escaping}, given a rate-of-change threshold \(L>0\), the sharp-variation region \(D_L\) is defined as
\begin{equation}
    \label{eq: weak continuous def}
    D_L\coloneqq\{\boldsymbol{x}\in \mathcal{X}\colon \exists \boldsymbol{x}'\in \mathcal{X}\,|f(\boldsymbol{x}) - f(\boldsymbol{x}')| >L \|\boldsymbol{x} - \boldsymbol{x}'\| \}.
\end{equation}
In this definition, $D_L$ contains points from which the function value changes more rapidly than the threshold \(L\). While Equation \eqref{eq: weak continuous def} provides a precise definition, it is impractical for numerical evaluation, as it requires evaluating all pairs of points.
For implementation, we use the first-order local approximation $f(\boldsymbol{x}')\approx f(\boldsymbol{x})+\nabla f(\boldsymbol{x})^\top(\boldsymbol{x}^\prime-\boldsymbol{x})$. This leads to the approximation
\(
  D_L\approx\{\boldsymbol{x}\in \mathcal{X}\colon
  \|\nabla f(\boldsymbol{x})\|>L\}.
\) This gradient-norm representation is a computational proxy for the rate-of-change definition in \eqref{eq: weak continuous def} and is evaluated through the fitted emulator. For nonsmooth simulators, the DGP gradient is used as an indicator of rapid change rather than as a true simulator gradient, which may not exist. In implementation, we represent the sharp-variation region as a super-level set of \(\|\nabla f(\boldsymbol{x})\|\) at the threshold \(L\). Examples of \(D_L\) are shown in Figure~\ref{fig:high_variation_region}. Based on this representation, we propose a sequential design criterion named \textbf{GradEnt}. GradEnt adapts entropy-based LSE strategies~\cite{gotovos2013active} to guide sampling toward locations where the estimated sharp-variation region is uncertain while also accounting for global predictive uncertainty.

\subsection{Acquisition criterion}
\label{subsec: Acquisition criterion}
In sequential design, an emulator is first built from an initial space-filling design $\mathcal{D}_0=\{\boldsymbol{x}_i,y_i\}_{i=1}^{n_0}$, commonly chosen via LHD \cite{santner2003design}. The process then iterates over the remaining budget $n-n_0$. At step $(i+1)$:
(a) the emulator is used to optimize an acquisition criterion $J_i$, yielding the next evaluation point $\boldsymbol{x}_{i+1}$;
(b) the design set is updated as $\mathcal{D}_{i+1}=\mathcal{D}_i\cup\{\boldsymbol{x}_{i+1},y_{i+1}\}$, and the emulator is retrained. Formally, the next design point is chosen by
\begin{equation}
    \boldsymbol{x}_{i+1}=\arg\max_{\boldsymbol{x}\in \mathcal{X}_{\text{cand}}}J_i(\boldsymbol{x}),
    \label{eq:general_acq_crt}
\end{equation}
where $\mathcal{X}_{\text{cand}}=\{\boldsymbol{x}_1,...,\boldsymbol{x}_{N_{\text{cand}}}\}$ is the candidate set with size $N_{\text{cand}}$, which represents a discretized version of the input space. Common choices for $J$ are variance-based criteria such as ALM, ALC, and MI \cite{krause2008near,beck2016sequential}; see Garud et al. (2017)~\cite{garud2017design} for a review. The variance-based criteria primarily measure uncertainty in the response value, but they do not directly quantify the local rate of change of the response, which is central to our target functions.

We approximate $D_L$ in \eqref{eq: weak continuous def} by a super-level set of the gradient norm. This reformulates identifying $D_L$ as an LSE problem. For LSE, a widely used sequential design criterion is the entropy-based method proposed by Gotovos et al.~\cite{gotovos2013active}. Denote $p_{\boldsymbol{x}}=p\text{\large (}\|\nabla f(\boldsymbol{x})\|\geq L\text{\large)}$ as the probability of $\boldsymbol{x}\in D_L$; the entropy criterion is then defined as
\begin{equation}
    \label{eq:entropy_criterion}
    J_{\text{ent}}(\boldsymbol{x})=-p_{\boldsymbol{x}}\log{(p_{\boldsymbol{x}})}-(1-p_{\boldsymbol{x}})\log{(1-p_{\boldsymbol{x}})}\,.
\end{equation}
The entropy-based acquisition $J_{\text{ent}}(\boldsymbol{x})$ is maximized at $p_{\boldsymbol{x}}\approx0.5$, i.e., along the decision boundary, where the sharp-region classification is most uncertain \cite{bect2012sequential,gotovos2013active}. Consequently, we sample near this frontier rather than in regions with $p_{\boldsymbol{x}}\approx 0$ or \(1\), maximizing expected entropy reduction and avoiding redundant evaluations. As the design proceeds, the emulator and $D_L$ are updated iteratively. The $p_{\boldsymbol{x}}=0.5$ contours (see Figure~\ref{fig: gradient norm surface}) contract toward the true transition surfaces, sharpening the estimated high-variation region (Section~\ref{Sec:experiment}).

\paragraph{Distribution on gradient norm} In the standard LSE problem, $p_{\boldsymbol{x}}$ can be evaluated through the cumulative distribution function (CDF) of the predictive distribution of the GP emulator. However, our focus is on the super-level set of the gradient norm rather than on the function value itself. This requires estimating the distribution of the norm of random Gaussian vectors. The following proposition enables analytical evaluation $p_{\boldsymbol{x}}$ when $\nabla f$ follows a Gaussian distribution.

\begin{proposition}
Let $\boldsymbol{x}\in\mathbb{R}^d$ and suppose the distribution of the gradient is \(
\nabla f(\boldsymbol{x}) \sim 
\mathcal{N}\bigl(\mu_\nabla(\boldsymbol{x}),\,\Sigma_\nabla(\boldsymbol{x})\bigr).
\) Then, $\|\nabla f(\boldsymbol{x})\|^2$ follows a generalized noncentral chi-squared distribution with CDF \(F_{\mathrm{G}\chi^2}\).
Consequently, for any threshold $L > 0$,
\[
p_{\boldsymbol{x}} := p\bigl(\|\nabla f(\boldsymbol{x})\|\ge L\bigr)
= p\bigl(\|\nabla f(\boldsymbol{x})\|^2\ge L^2\bigr) = 1 - F_{\mathrm{G}\chi^2}\bigl(L^2;\,\boldsymbol{w}(\boldsymbol{x}),\,\boldsymbol{\delta}(\boldsymbol{x})\bigr), 
\]
where the weights $\boldsymbol{w}(\boldsymbol{x})$ and noncentrality parameters $\boldsymbol{\delta}(\boldsymbol{x})$ are determined by the spectral decomposition of $\Sigma_\nabla(\boldsymbol{x})$. Specifically, if $\{\lambda_i(\boldsymbol{x}), \boldsymbol{u}_i(\boldsymbol{x})\}_{i=1}^d$ are the eigenvalue-eigenvector pairs of $\Sigma_\nabla(\boldsymbol{x})$, then:
\(
w_i(\boldsymbol{x}) = \lambda_i(\boldsymbol{x})\) and \(
\delta_i(\boldsymbol{x}) = \frac{\boldsymbol{u}_i(\boldsymbol{x})^\top \mu_\nabla(\boldsymbol{x})}{\sqrt{\lambda_i(\boldsymbol{x})}}.
\)
\end{proposition}
This proposition provides an exact evaluation of $p_{\boldsymbol{x}}$ in terms of the Gaussian predictive distribution. However, evaluating the CDF $F_{\mathrm{G}\chi^2}$ of a generalized noncentral chi-square
distribution typically requires specialized numerical algorithms (e.g., Imhof- or
Davies-type methods; see \cite{das2025new,das2021method,imhof1961computing}). To obtain an efficient approximation, we
replace $\Sigma_\nabla(x)$ with an \textit{isotropic}\footnote{An anisotropic diagonal covariance would still lead to a weighted sum of noncentral chi-square random variables, and hence require evaluating a generalized noncentral chi-square distribution.} covariance, \(
s^2(\boldsymbol{x})\coloneqq\frac{1}{d}\sum_{i=1}^d\sigma^2_{\nabla,i}(\boldsymbol{x}) = \frac{1}{d}\,\mathrm{tr}\big(\Sigma_\nabla(\boldsymbol{x})\big),
\) and approximate the gradient distribution as 
\( \mathcal N\bigl(\mu_\nabla(\boldsymbol{x}),\,s^2(\boldsymbol{x})\,I_d\bigr).
\) This approximation preserves the overall scale of gradient uncertainty while enabling a tractable representation \cite{sullivan2015introduction}. Then, the scaled norm satisfies
\(
\frac{\|\nabla f(\boldsymbol{x})\|}{s(\boldsymbol{x})} \sim \chi\bigl(d,\lambda_\chi(\boldsymbol{x})\bigr).
\)
Hence,
\(
p_{\boldsymbol{x}} \approx 1 - F_{\chi}\!\Bigl(\tfrac{L}{s(\boldsymbol{x})};\,d,\,\lambda_\chi(\boldsymbol{x})\Bigr),
\) where $F_{\chi}$ denotes the CDF of the noncentral chi distribution with degrees of freedom $d$ (i.e., the dimension of input $\boldsymbol{x}$) and the noncentrality parameter $\lambda_\chi(\boldsymbol{x})=\sqrt{\sum_{i=1}^d({\mu_\nabla}_i\big(\boldsymbol{x})/s(\boldsymbol{x})\big)^2}$. 

\subsection{Practical consideration}
\label{subsec: practical consideration}
\paragraph{Adaptive threshold} In GradEnt, \(L\) is used as a design threshold that guides sampling toward regions of relatively large local variation, and is progressively updated adaptively as the emulator is refined. For the true function $f$, one can show that $L=\max_{\boldsymbol{x}\in \mathcal{X}\setminus D}\|\nabla f(\boldsymbol{x})\|$ is a valid Lipschitz constant in the domain $\mathcal{X}\setminus D$ \cite{gonzalez2016batch}. We approximate it using
\begin{equation}
\label{eq: repulsion term}
    L_{\text{DGP}}=\max_{\boldsymbol{x}\in \mathcal{X}\setminus D}\|\mu_{\nabla,\text{DGP}}(\boldsymbol{x})\|.
\end{equation}
In practice, $L_{\text{DGP}}$ is updated iteratively. At iteration $i$, it is set to the maximum gradient norm over the \textit{pruned} candidate set
$\mathcal{X}^{-}_{\text{cand}}=\{\boldsymbol{x}\in\mathcal{X}_{\text{cand}}|p_{\boldsymbol{x}}<\delta,L=L_{\text{DGP}}^{i-1}\}$, where \(\delta\in(0,1)\) is the classification threshold for the probability that \(\boldsymbol{x}\) belongs to \(D\). The update therefore excludes candidates already likely to be in the sharp-variation region and takes the maximum predicted gradient norm over the remaining candidates. Small $\delta$ overestimates the sharp-variation region $D$ (larger $D$, fewer false negatives), whereas large $\delta$ underestimates $D$ (fewer false positives).  In the first iteration, the full candidate set is used to compute $L_{\text{DGP}}^0$.

\paragraph{Over concentration} Sharp variations can induce very large gradients, causing the acquisition function to over-concentrate samples on a small portion of the transition hypersurface rather than covering it well. To address this issue, we formulate GradEnt acquisition as a bi-objective optimization on the Pareto frontier \cite{ngatchou2005pareto}, jointly trading off entropy $J_{\text{ent}}$ (exploitation) and predictive uncertainty $\sigma_{\text{DGP}}^2$ (exploration), adapting the Pareto-based strategy of Sauer et al. \cite{sauer2024actively} for contour estimation. We use predictive uncertainty for exploration to prevent over-concentration. Although gradient uncertainty is also a valid choice for exploration, it is inherently included in the exploitation term $J_{\text{ent}}$. Formally, we define the acquisition criteria as a two-dimensional vector \(\boldsymbol{J}(\boldsymbol{x})=\big(J_{\text{ent}}(\boldsymbol{x}),\sigma_{\text{DGP}}^2(\boldsymbol{x})\big)\) aimed at maximizing both objectives. We denote $\boldsymbol{J}(\boldsymbol{x}_i)\succ\boldsymbol{J}(\boldsymbol{x}_j)$ if $\boldsymbol{J}(\boldsymbol{x}_i)$ \textit{strictly dominates} $\boldsymbol{J}(\boldsymbol{x}_j)$, meaning that $J_{\text{ent}}(\boldsymbol{x}_i)>J_{\text{ent}}(\boldsymbol{x}_j)$ and $\sigma_{\text{DGP}}^2(\boldsymbol{x}_i)>\sigma_{\text{DGP}}^2(\boldsymbol{x}_j)$. The Pareto frontier is then defined as \cite{ishizaka2013multi}:
\begin{equation*}
    \mathcal{P}(\mathcal{X})=\big\{\boldsymbol{x}\in\mathcal{X}\,|\,\{\boldsymbol{x}'\in\mathcal{X}\,|\,\boldsymbol{J}(\boldsymbol{x}')\succ\boldsymbol{J}(\boldsymbol{x}),\boldsymbol{x}'\neq\boldsymbol{x}\}=\emptyset\big\}.
\end{equation*}
It consists of all non-dominated points, where no other candidate improves both criteria simultaneously. In sequential design, we approximate $\mathcal{P}(\mathcal{X})$ over the discretized candidate set $\mathcal{X}_{\text{cand}}$. The frontier is obtained efficiently by: (1) sorting candidates in descending order of one criterion (ensuring maxima are always included), and (2) traversing the list, adding a candidate to $\mathcal{P}(\mathcal{X}_{\text{cand}})$ if it exceeds all previous points in the second criterion. Finally, the next design point is drawn uniformly from $\mathcal{P}(\mathcal{X}_{\text{cand}})$. In our problem, the two objectives have different scales, and we do not impose a fixed preference between them. Uniform sampling provides an efficient selection rule. Alternative advances, such as Expected Hypervolume Improvement \citep{emmerich2011hypervolume,yang2019efficient}, are promising and may further improve the choice from the Pareto frontier.

In summary, Algorithm \ref{alg: Gradent-R} outlines the proposed GradEnt method.
\begin{algorithm}
\caption{GradEnt}
\label{alg: Gradent-R}
\begin{algorithmic}[1]
\REQUIRE number of designs $N$; computer model $f$; classification threshold $\delta$
\STATE Select initial design $\mathcal{D}_0=\{\boldsymbol{x}_i,f(\boldsymbol{x}_i)\}_{i=1}^{n_0}$
\STATE Initialize DGP emulator with $\mathcal{D}_0$
\FOR{$t \gets 0$ to $(N - n_0 -1)$}
    \STATE Select the candidate set $\mathcal{X}_{\text{cand}}\subset \mathcal{X}$
    \STATE Compute $\{\mu_{\text{DGP}}(\boldsymbol{x}),\sigma_{\text{DGP}}^2(\boldsymbol{x})|\boldsymbol{x}\in\mathcal{X}_{\text{cand}}\}$ with Equation \eqref{eq: DGP predicitive distribution}.
    \STATE Determine $\{\mu_{\nabla,_{\text{DGP}}}(\boldsymbol{x}),\Sigma_{\nabla,_{\text{DGP}}}(\boldsymbol{x})|\boldsymbol{x}\in\mathcal{X}_{\text{cand}}\}$ with Equation \eqref{eq: DGP gradient covariance}.
    \STATE Compute pruned set \\
    $\mathcal{X}^{-}_{\text{cand}}=\{\boldsymbol{x}\in\mathcal{X}_{\text{cand}}|p_{\boldsymbol{x}}<\delta,L=L_{\text{DGP}}^{t-1}\}$ if $t>0$ else $\mathcal{X}^{-}_{\text{cand}}=\mathcal{X}_{\text{cand}}$
    \STATE Compute threshold $L^t_{\text{DGP}}=\max_{\boldsymbol{x}\in\mathcal{X}^{-}_{\text{cand}}}\|\mu_{\nabla,_{\text{DGP}}}(\boldsymbol{x})\|$
    \STATE Compute $\boldsymbol{J}(\boldsymbol{x})=\big(J_{\text{ent}}(\boldsymbol{x}),\sigma_{\text{DGP}}^2(\boldsymbol{x})\big)$ with Pareto frontier $\mathcal{P}(\mathcal{X}_{\text{cand}})$.
    \STATE Select next design point $\boldsymbol{x}_*\sim\mathcal{P}(\mathcal{X}_{\text{cand}})$ 
    \STATE Run computer model $f(\boldsymbol{x}_*)$
    \STATE Update design set $\mathcal{D}_{t+1}\leftarrow\mathcal{D}_{t}\cup\{\boldsymbol{x}_*, f(\boldsymbol{x}_*)\}$ and DGP emulator with $\mathcal{D}_{t+1}$
\ENDFOR
\end{algorithmic}
\end{algorithm}

\section{Experiment}
\label{Sec:experiment}
In this section, we validate our proposed sequential design in a series of synthetic simulations and a real-world application. We perform a comparison with the following design methods and emulators: 
\begin{itemize}
    \item Active Learning MacKay with GP and TGP (ALM-GP, ALM-TGP) \cite{mackay1992information};
    \item Mutual Information for Computer Experiments with GP (MICE-GP) \cite{beck2016sequential};
    \item Variance of Improvement for Global Fit with GP (VIGF-GP) \cite{mohammadi2022sequential};
    \item Active Learning Cohn with DGP and TGP (ALC-DGP, ALC-TGP) \cite{gramacy2008bayesian,sauer2023active}.
\end{itemize}
These include standard GP-based criteria (ALM-GP, MICE-GP, VIGF-GP) and nonstationary emulators (ALC-DGP, ALM/ALC-TGP). We also include GradEnt-GP, which denotes GradEnt with a GP emulator and is introduced to isolate the effect of the acquisition criterion from that of the DGP emulator. 

\paragraph{Implementation} We implement ALC-DGP using the R package \texttt{deepgp}
\cite{sauer2023active} and the two TGP-based methods using the R package
\texttt{tgp} \cite{gramacy2007tgp}. For ALM-GP, MICE-GP, VIGF-GP, and GradEnt, we use the same version of
the Python package \texttt{dgpsi}, which includes our gradient extension\footnote{The gradient implementation is available
as the GitHub release 
\href{https://github.com/yyimingucl/DGP/releases/tag/v2.5.0-grad1}
{\texttt{dgpsi v2.5.0-grad1}}. The experimental code for sequential design is provided in the SM.}. We set the classification threshold $\delta=0.75$ and report a sensitivity analysis for $\delta$ at the end of this section. All emulators use SE kernels. DGPs are specified as two-layer models with two GP nodes in the first layer.

\paragraph{Evaluation}
We use four metrics: global/local Normalized Root Mean Square Error Percent (NRMSEP) and global/local Normalized Continuous Ranked Probability Score ($\text{NCRPS}_y$). Global metrics evaluate performance over $\mathcal{X}$, while local metrics focus on the sharp-variation region $D$. NRMSEP is defined as
\begin{equation*}
    \text{NRMSEP}
    =\frac{100\sqrt{\frac{1}{n}\sum_{i=1}^{n}(\mu(\boldsymbol{x}_i)-f(\boldsymbol{x}_i))^2}}
           {\max\big(\{f(\boldsymbol{x}_i)\}_{i=1}^n\big)-\min\big(\{f(\boldsymbol{x}_i)\}_{i=1}^n\big)},
\end{equation*}
where $f(\boldsymbol{x}_i)$ is the computer model output and $\mu(\boldsymbol{x}_i)$ is the emulator mean. $\operatorname{NCRPS}_{y}$ is the scalar-output ($d=1$) version of
\Cref{eq:ncrps-gradient}, applied to the predictive distribution of the emulator output. Global metrics are computed on an LHD test set over $\mathcal{X}$. For local metrics, test points are generated near known transition surfaces by sampling an LHD over the free surface coordinates, solving for constrained coordinates, and adding small Gaussian perturbations. Points that fall outside the input domain after perturbation are rejected and resampled. Table~\ref{tab: sequential_design_settings} summarizes the main sequential design settings. Further details on emulator settings, candidate-set construction, and test-set generation are provided in SM Section~4.

\begin{table}[ht]
      \centering
      \caption{Sequential design settings used in the main experiments. Candidate sizes are reported as GP-based / nonstationary GP-based sequential designs.}
      \label{tab: sequential_design_settings}
      \small
      \begin{tabular}{lcccc}
          \toprule
          Experiment & Initial size & Final size & Candidate size & Replicates\\
          \midrule
          Plateau, \(d=2\) & 5 & 100 & 500 / 200 & 40 \\
          Plateau, \(d=5\) & 25 & 300 & 1000 / 300 & 40 \\
          Plateau, \(d=8\) & 50 & 500 & 1500 / 500 & 10 \\
          Plateau multiple, \(d=2\) & 5 & 100 & 500 / 200 & 40 \\
          RB convection, \(d=2\) & 10 & 200 & 500 / 200 & 40 \\
          \bottomrule
      \end{tabular}%
      
      \vspace{-0.3em}
\end{table}

\subsection{Synthetic simulation}
We first consider a synthetic computer model with $d$ dimensional input $\boldsymbol{x}\in[-1,1]^d$:
\begin{equation*}
    \label{eq: Plateau function}
    f(\boldsymbol{x})=2\Phi\Big(\sqrt{2}(-4-\alpha\sum_{i=1}^dx_i)\Big)-1.
\end{equation*}
which is introduced in \cite{izzaturrahman2022modeling} as a test case for localizing the fast variations and contour finding. The coefficient \(\alpha\) controls the steepness and is set to \(24\) in all experiments. The hyper-surface $\sum_{i=1}^dx_i=-\frac{4}{\alpha}$ is the transition surface. We first study the low-dimensional case \(d=2\), and then examine \(d=5\) and \(d=8\) to assess higher-dimensional performance.

\begin{figure}
    \centering
    \includegraphics[width=0.95\linewidth]{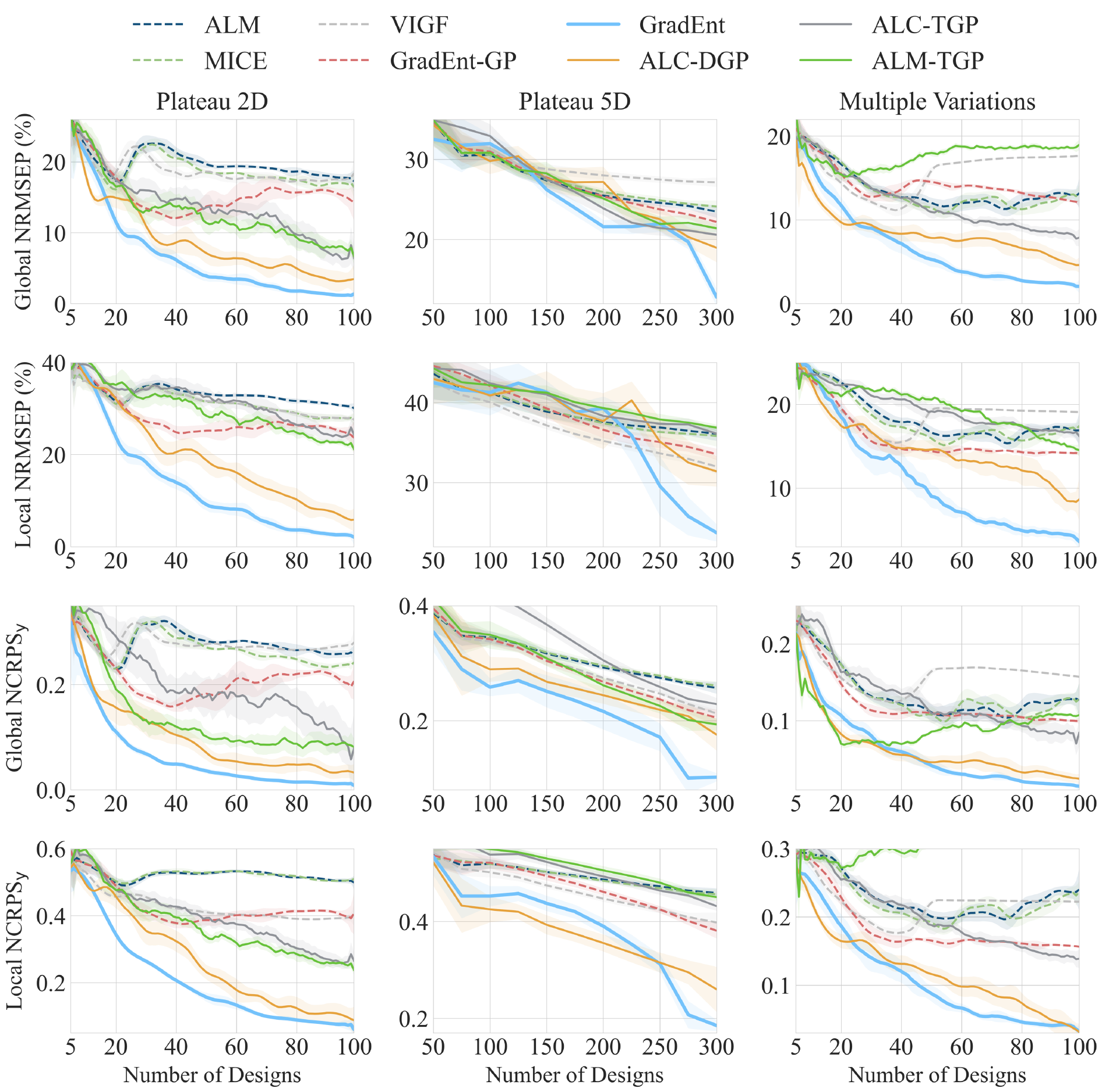}
    \vspace{-1.em}
    \caption{Sequential design performance on three plateau-function benchmarks. Dashed curves correspond to stationary GP emulators and solid curves to nonstationary GP emulators. Each column represents a test problem and each row a performance metric. Curves show the mean performance over 40 runs, with shaded areas for standard deviations.}
    \label{fig:all_seq_design_results}
    \vspace{-1.4em}
\end{figure}
\paragraph{$\boldsymbol{d}\boldsymbol{=}\boldsymbol{2}$}

The first column of Figure~\ref{fig:all_seq_design_results} reports the global and local NRMSEP and $\text{NCRPS}_y$ over iterations. The strong nonstationarity makes the transition region challenging for GP-based methods, and their global and local errors tend to plateau after roughly 40 samples. TGP-based methods improve over the GP baselines by introducing local flexibility, but their partition-based structure is less stable than the DGP emulator in this example. For ALC-DGP, NRMSEP and $\text{NCRPS}_y$ converge to reasonable values but with larger variability across runs, whereas GradEnt converges more stably with consistently lower global and local errors. Figure~\ref{fig: gradient norm surface} (left) further shows that the estimated probability \(p_{\boldsymbol{x}}\) is high near the sharp-variation region

\paragraph{$\boldsymbol{d}\boldsymbol{=}\boldsymbol{5}\boldsymbol{,}\boldsymbol{8}$} 
Metrics are recorded every 25 iterations for \(d=5\) and every 90 iterations for \(d=8\). The second column of
Figure~\ref{fig:all_seq_design_results} shows the \(d=5\) results, while
the corresponding \(d=8\) results are provided in Figure~SM7 of the SM. In both cases, GradEnt continues to reduce both global and local errors as the number of designs grows, with particularly strong gains in local metrics relative to other methods.
ALC-DGP also improves upon the GP- and TGP-based methods, but its performance is less stable and generally remains above that of GradEnt. For GP-based methods, performance decreases as the dimension increases; by \(d=8\), GradEnt-GP performs similarly to the other GP-based designs. 

\paragraph{Multiple Variations} To assess performance in the presence of multiple sharp variation regions, we modify the plateau function as follows:
\begin{equation*}
    \label{eq:two_sharp_plateau}
    f(\boldsymbol{x}) = \frac{3}{4}\left[2
        \Phi\Big(\sqrt{2}(h - \alpha\sum_{i=1}^d x_i)\Big) - 1\right ]
        \;+\;
        \frac{1}{4}\left[2\Phi\Big(\sqrt{2}(-h - \alpha\sum_{i=1}^d x_i)\Big) - 1\right ],
\end{equation*}
where $h > 0$ specifies the separation between the two sharp variations. We set $h = 15$ and retain $\alpha = 24$ as in the single-boundary case. The steep transitions occur along the two hyperplanes:
\(
\sum_{i=1}^d x_i = -\frac{h}{\alpha}\)
and \(
\sum_{i=1}^d x_i = \frac{h}{\alpha},
\)
with the plateau region lying between them. This formulation introduces two nonstationary transitions with unequal steepness, providing a more challenging test for local and global emulations. Figure~\ref{fig:all_seq_design_results} (third column) shows that GradEnt outperforms the alternatives across all metrics and exhibits stable loss curves. While ALC-DGP remains competitive in probabilistic metrics, it shows higher predictive errors than GradEnt. As shown in Figure~\ref{fig: gradient norm surface} (mid), the estimated probability surface has large values near the multiple variation regions. We also observe a localized high-probability region in the bottom-left corner, which is attributable to boundary effects from sparse sampling in that area, as shown in Figure~SM9 of the SM.

\subsection{Flow Phase Transition in Rayleigh-B\'enard Convection}
\label{subsec: flow-transistion}
\begin{figure}
    \centering    
    \includegraphics[width=0.73\linewidth]{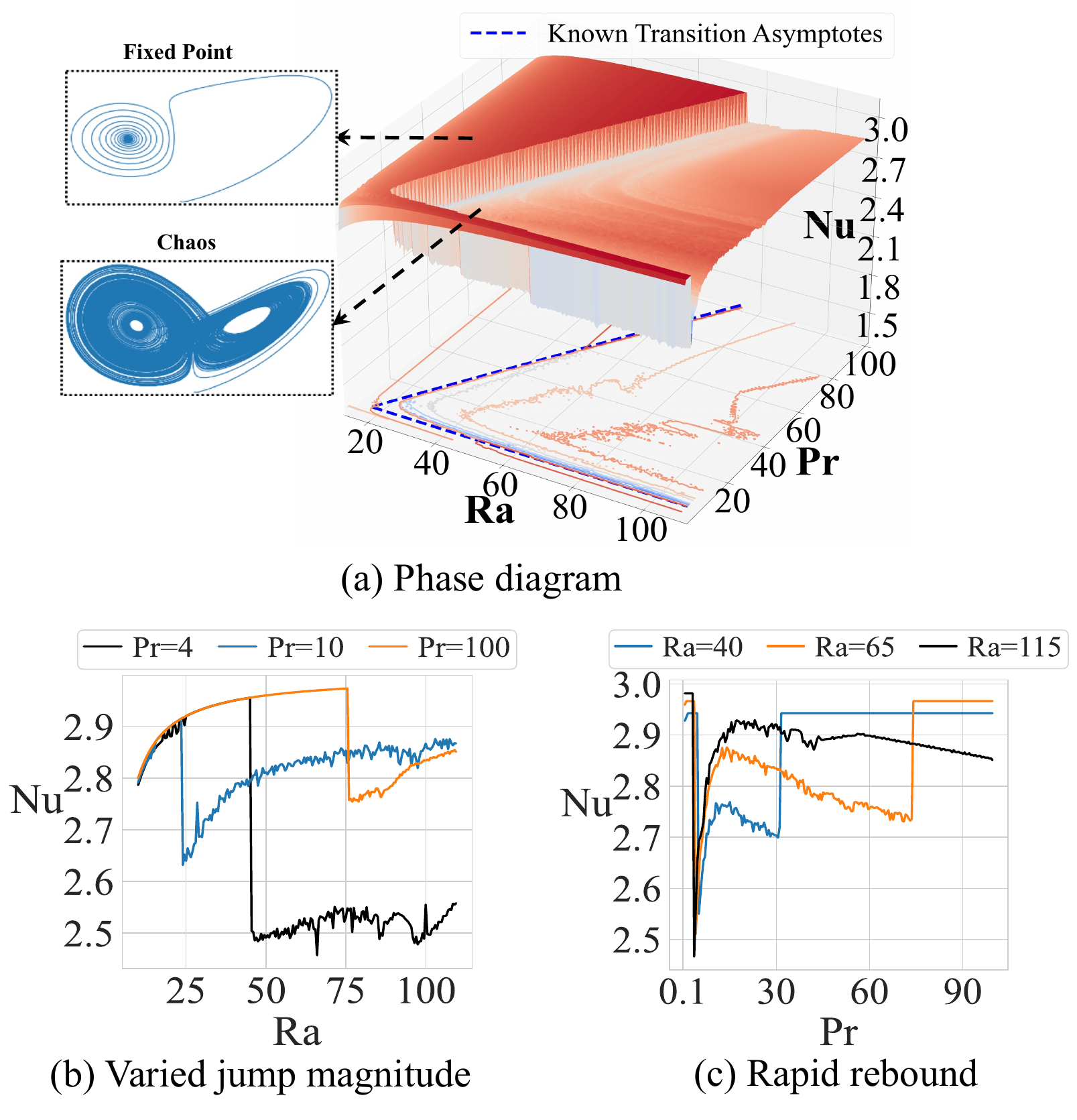}
    \vspace{-0.75em}
    \caption{(a) The $(Ra, Pr) - Nu$ surface of the Lorenz-63 model. The dashed blue lines represent known transition asymptotes that induce instability in the Lorenz-63 system. Further details can be found in \cite{dullin2007extended}. (b) Cross-sections of the phase diagram at $Pr = 4$, $10$, and $100$ illustrate the varying jump magnitudes along the transition boundary. (c) Cross-sections of the phase diagram at $Ra = 40$, $65$, and $115$ highlight the rapid rebound along the transition boundary.}
    \label{fig:ra-pr-nu surface}
    \vspace{-1em}
\end{figure}
\begin{figure}
    \centering
    \includegraphics[width=0.8\linewidth]{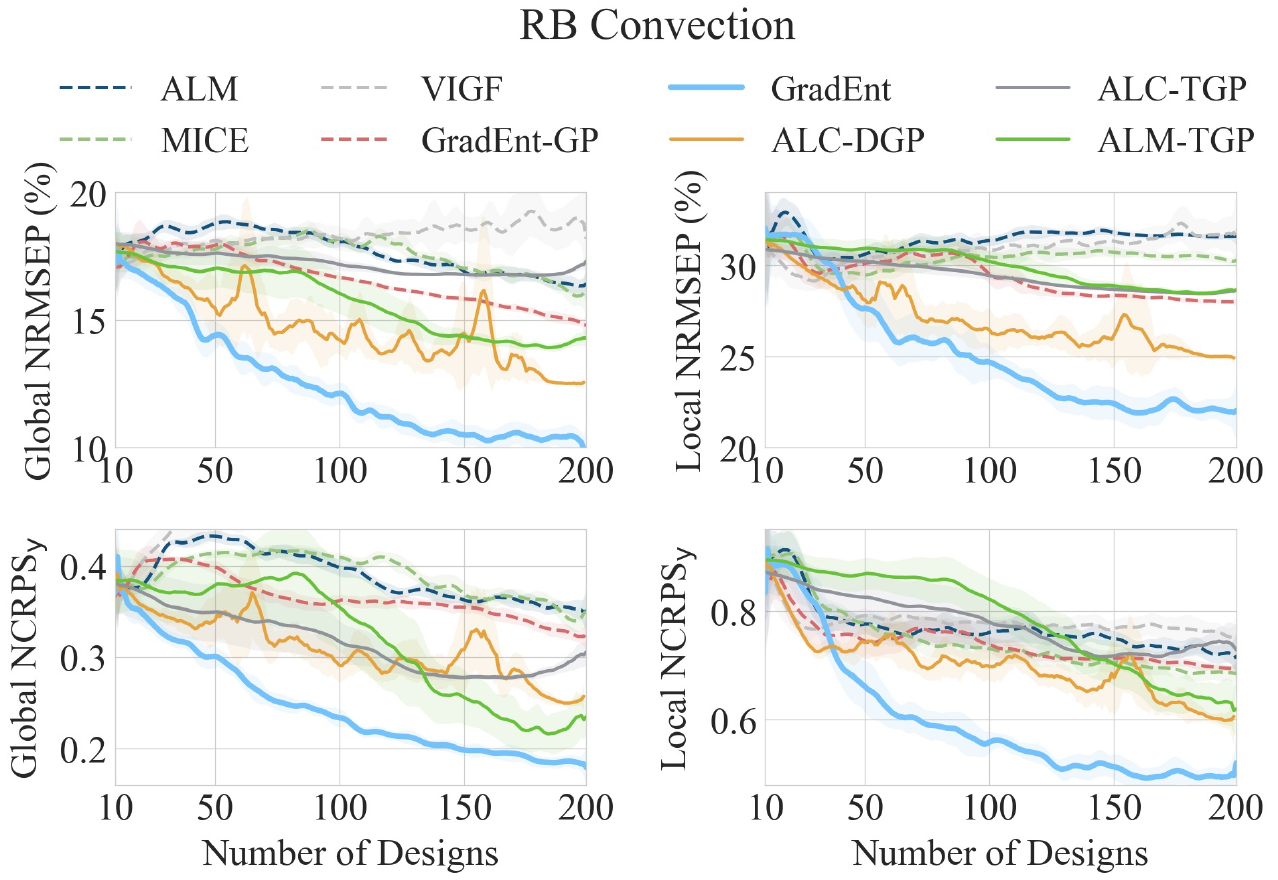}
    \vspace{-0.7em}
    \caption{Sequential design performance on the real-world RB convection benchmark. Panels show four performance metrics; dashed and solid curves denote stationary and nonstationary GP emulators, respectively. Shading indicates standard deviations over 40 runs.}
    \label{fig:seq_design_results_rb_convection}
    \vspace{-0.7em}
\end{figure}
\begin{figure}
    \centering
    \includegraphics[width=0.9\linewidth]{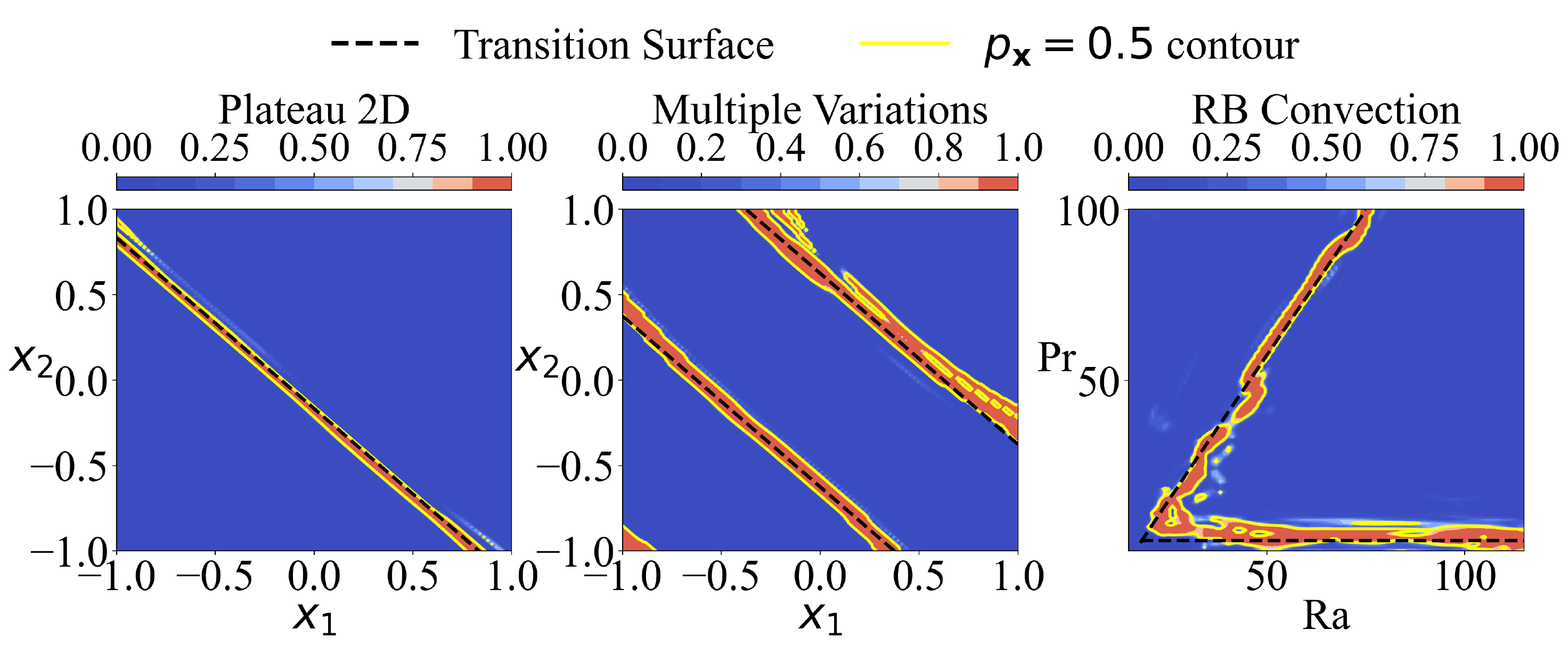}
    \vspace{-0.7em}
    \caption{Visualization of the estimated probability surfaces for Plateau 2D (left), multiple variations (middle), and RB convection (right). For Plateau 2D and
    multiple variations, \(p_{\boldsymbol{x}}\) is evaluated using \(L=36\) and \(L=15\), respectively; for RB convection, \(L=89.2\) is estimated from the
    final sequential design iteration. The yellow solid line denotes the \(p_{\boldsymbol{x}}=0.5\) contour, and the black dashed line denotes the known
    transition surface.}
    \label{fig: gradient norm surface}
    \vspace{-0.8em}
\end{figure}
The Lorenz-63 model \cite{lorenz1963deterministic}, which consists of three coupled nonlinear ODEs, 
\begin{equation*}
\frac{dx}{dt}=\sigma(y-x),\,\,\frac{dy}{dt}=x(\rho-z)-y,\,\,\frac{dz}{dt}=xy-\beta z,
\end{equation*} 
was originally introduced as a simplified model of Rayleigh--B\'enard convection: an incompressible fluid between two plates perpendicular to the earth’s gravitational force. The parameters $\sigma$, $\rho$, and $\beta$ represent the Prandtl (Pr) number, the Rayleigh (Ra) number, and the coupling strength, respectively, with $\beta$ set to $\frac{8}{3}$. In convective flow, the Nusselt number (Nu) quantifies heat transfer efficiency and identifies laminar and chaotic regimes with bifurcations. Understanding the transition from stable (laminar) to chaotic (turbulent) flow is critical in fields like atmospheric science for weather modeling \cite{lorenz1963deterministic} and engineering for designing systems like heat exchangers and electronic cooling solutions \cite{ahlers2009heat}. The Lorenz-63 model analytically characterizes sharp behavioral shifts, making it an important benchmark for emulating abrupt transitions.

In the Lorenz–63 model, it is estimated as $\text{Nu} = 2\frac{z_{\infty} - z_0}{\rho}$ \cite{robbins1979periodic}, where fixed points correspond to laminar flow and chaotic attractors correspond to turbulent flow. We estimate \(z_{\infty}\) by integrating the system with a time step of \(\Delta t=10^{-3}\) for \(2\times 10^5\) steps, discarding the first half as burn-in and averaging over the remaining trajectory. A sensitivity of $z_\infty$ to the integration length is reported in SM Section 4.1.3. The locations of the phase transitions are given analytically, as shown in Figure \ref{fig:ra-pr-nu surface}(a); see \cite{dullin2007extended} for more details. We consider the parameters (Ra, Pr) within the domain $[15,115] \times [0.1,100.1]$ as inputs to the model, with Nu as the output. Two challenges arise: (1) jump magnitudes vary along transition boundaries, with larger jumps being easier to capture than smaller ones (Figure~\ref{fig:ra-pr-nu surface} (b)); (2) the surface exhibits sharp drops followed by rapid rebounds, which emulators may overshoot  (Figure~\ref{fig:ra-pr-nu surface} (c)) \cite{gorodetsky2014efficient}. 

\begin{figure}
    \centering
    \includegraphics[width=0.75\linewidth]{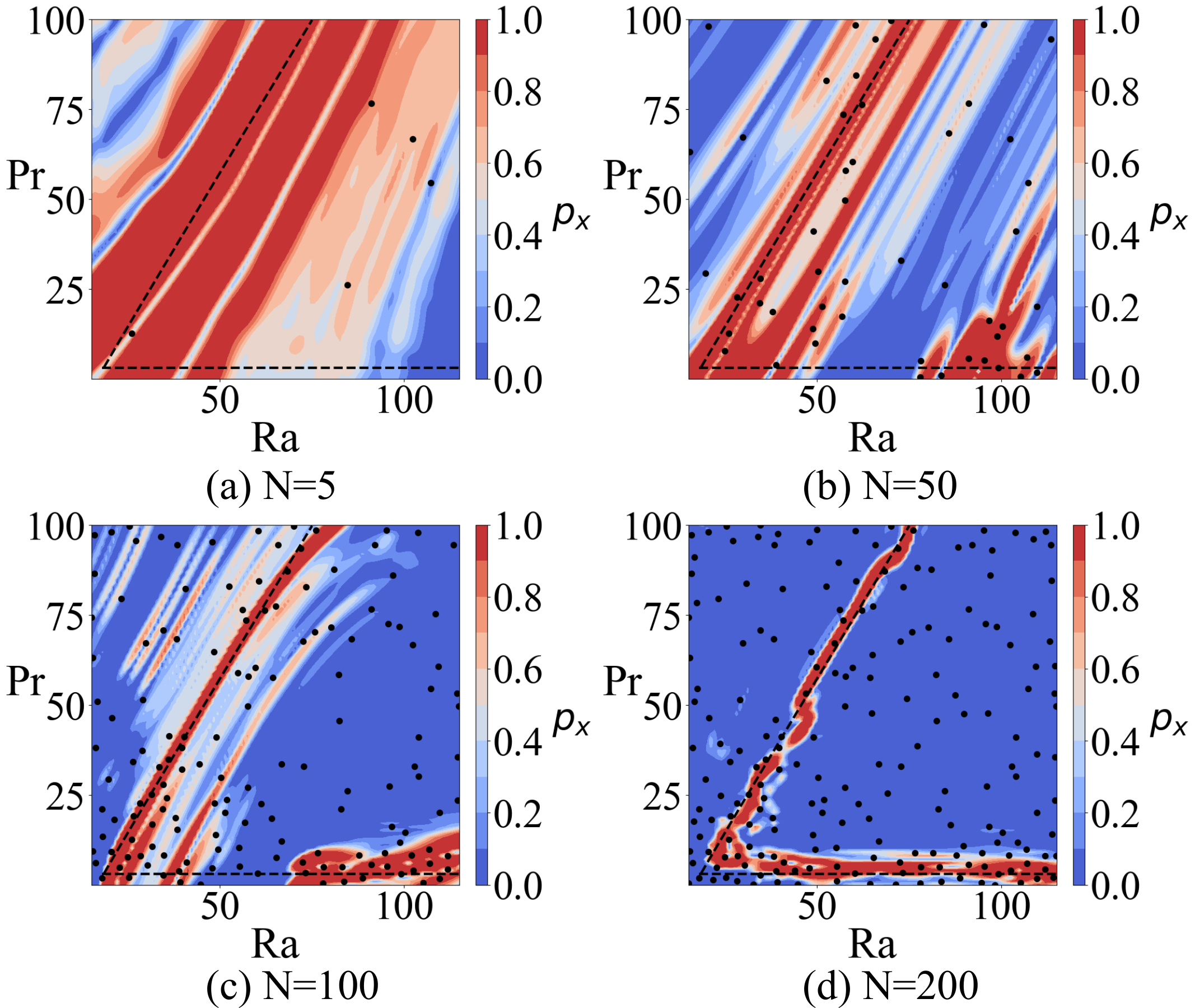}
    \vspace{-0.6em}
    \caption{Estimated probability surface $p_{\boldsymbol{x}}=p(\|\nabla f(\boldsymbol{x})\|>{L})$ for the RB convection problem (blue $\approx0$, red $\approx1$) across sequential design iterations. Black dots denote sampled design locations.}
    \label{fig: acq_evl_combined}
    \vspace{-1.2em}
\end{figure}

\paragraph{Results} Figure~\ref{fig:seq_design_results_rb_convection} shows that GradEnt achieves the lowest global and local NRMSEP and $\text{NCRPS}_y$ after about 50 samples and converges stably. The three GP-based methods (ALM, MICE, VIGF) struggle with the model’s nonstationarity, while ALC-DGP displays higher variance across replications. GradEnt prioritizes sampling near regions of sharp variation while maintaining global coverage through the Pareto-frontier objective, leading to improved global and local accuracy. In Figure~\ref{fig: gradient norm surface} (right), the contour plot visualizes the estimated probability $p_{\boldsymbol{x}}$ that $\boldsymbol{x}$ belongs to the sharp variation region $D$. This probability surface is computed using the constructed DGP emulator and the estimated threshold $L$ from the final iteration of the sequential design. Figure~\ref{fig: acq_evl_combined} further tracks this evolution over iterations: at $N=150,200$, $p_{\boldsymbol{x}}\approx1$ contracts into a narrow band that aligns with the known transition boundary.
\subsection{Sensitivity Analysis}
\begin{figure}[t]
    \centering \includegraphics[width=0.8\linewidth]{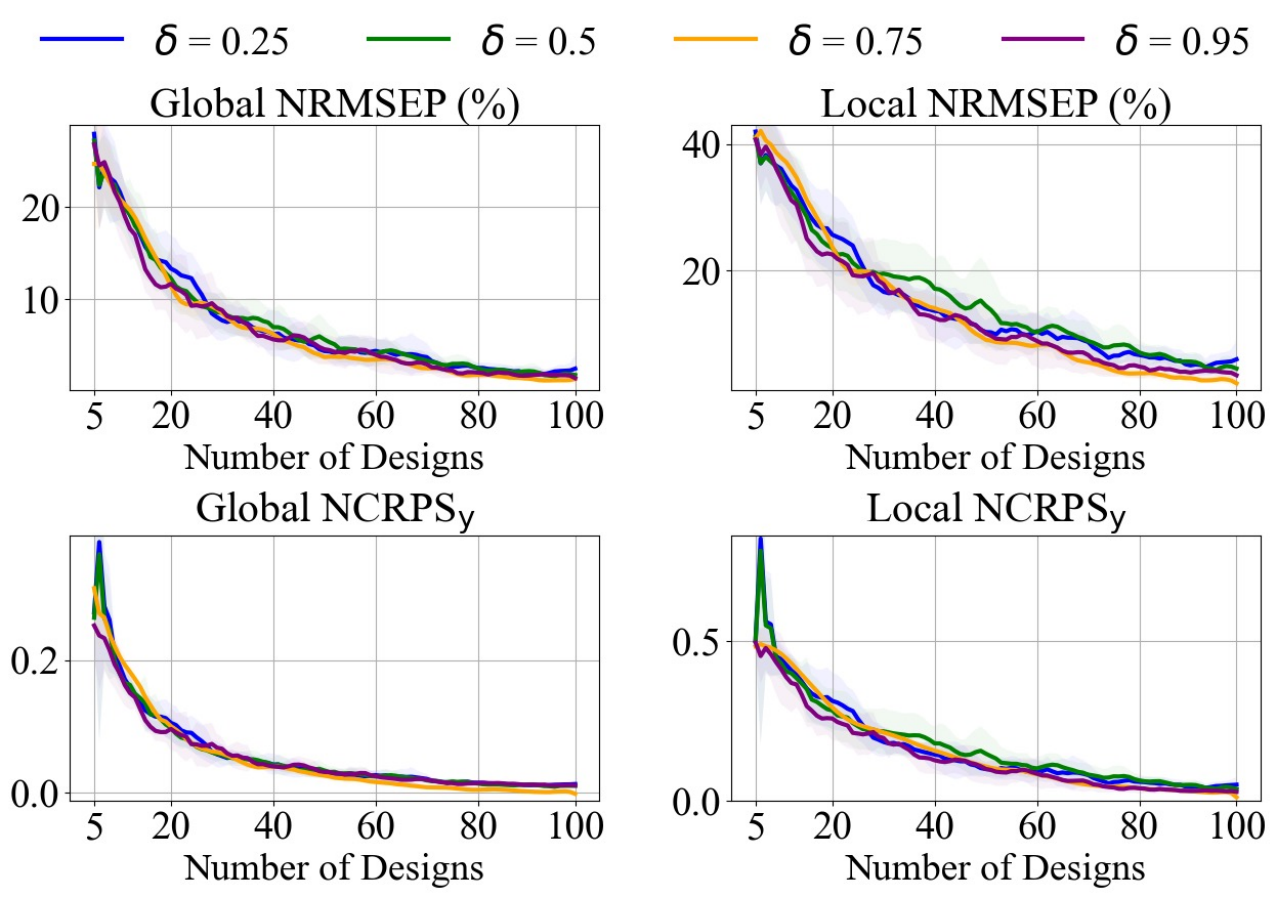}
    \vspace{-1em}
    \caption{Sensitivity analysis of $\delta$ on the 2D Plateau problem.}
    \label{fig: ablation_delta}
    \vspace{-1em}
\end{figure}

Figure~\ref{fig: ablation_delta} shows the sensitivity of GradEnt to the classification threshold $\delta$. We evaluate $\delta$ over $[0.25,0.5,0.75,0.95]$ on the 2D Plateau problem. The overall performance (Global NRMSEP and $\text{NCRPS}_y$) is robust to the classification threshold, with the performance curves being nearly identical. Local NRMSEP and $\text{NCRPS}_y$ improve for larger values ($\delta=0.75,0.95$), with a slight advantage at $\delta=0.75$. We also observe mild early-stage instability for smaller classification thresholds ($\delta=0.25,0.5$), which disappears as the design grows. With few initial samples, estimates are uncertain, and a small $\delta$ is more permissive, potentially overestimating the sharp variation region and inducing over-exploration. As additional data are collected, uncertainty decreases, and the estimate of \(L\) is refined, stabilizing both the estimated sharp-variation region and the emulation performance.

\section{Conclusion}
This paper develops an efficient approximation for the gradient uncertainty of DGP emulators. The proposed approach provides closed-form expressions for the gradient mean and covariance approximation, enabling gradient evaluation with UQ from function-value observations only. Experiments on nonstationary examples show improved gradient accuracy and efficient gradient UQ. The proposed DGP gradient approach may support applications, including sensitivity analysis and design optimization. Building on this gradient uncertainty, we develop GradEnt, a sequential design method for computer models with sharp variations. GradEnt views sharp-variation regions as super-level sets of the gradient norm and uses an entropy-based criterion to guide sampling toward uncertain sharp region estimates. This perspective also provides a probabilistic quantification of the estimated sharp-variation regions, which could be useful in safety-related applications. Across synthetic and real-world examples, GradEnt improves both global and local emulation accuracy relative to competing sequential design methods.

As guidance for practitioners, GradEnt is well suited when localized sharp variations, such as phase
transitions, are expected but their locations are unknown. When their presence is uncertain, the initial design can provide a preliminary diagnosis by assessing whether the posterior gradient norm is locally large relative to its uncertainty. As shown in SM Section~4.2.4, GradEnt remains competitive when such
variation is absent, a robustness that can be partly attributed to the inclusion of predictive variance as an exploration objective.

\paragraph{Limitations} This paper considers two-layer DGPs with squared exponential kernels,
which are analytically convenient and empirically effective. Extending
the proposed gradient approximation to deeper architectures and broader
kernel families remains future work. It should also be noted that, for nonsmooth simulators, the DGP
gradient should be viewed as an emulator-based indicator of rapid
local change rather than as a derivative estimate, since the true
derivative may not exist at nonsmooth points. Finally, GradEnt uses an entropy-based acquisition strategy; other strategies built on the proposed DGP gradient uncertainty may also be effective and represent valuable future directions.

\appendix
\section{Table of Notation}
\label{sec: notations}
\begin{center}
\small
\resizebox{0.99\textwidth}{!}{\begin{tabular}{c | c }
Notation & Meaning \\
 \hline
 \hline
 $x$ & 1-dimensional model input, $x\in\mathbb{R}$  \\ 
 
 $y$ & (global) model output, $y\in\mathbb{R}$ \\
 
 $\boldsymbol{x}$ & $d-$dimensional model input, $\boldsymbol{x}\in\mathbb{R}^d$ \\
 
 $x_{i,j}$ & $j$th dimension of $i$th sample, ${x_{i,j}}\in\mathbb{R}$  \\
 
 $\boldsymbol{y}$ & set of model outputs with size $m$,
 $\boldsymbol{y}\in\mathbb{R}^m$ \\
 $\boldsymbol{X}$ & set of $d-$dimensional model inputs with size $m$, $\boldsymbol{X}\in\mathbb{R}^{m\times d}$\\
 $\mu_{\text{GP}}$,$\mu_{\text{LGP}}$, $\mu_{\text{DGP}}$ & GP/LGP/DGP predictive mean \\ $\sigma^2_{\text{GP}}$,$\sigma^2_{\text{LGP}}$,$\sigma^2_{\text{DGP}}$ &  GP/LGP/DGP predictive variance \\
  $\mu_{\nabla,\text{GP}}$,$\mu_{\nabla,\text{LGP}}$, $\mu_{\nabla,\text{DGP}}$ & GP/LGP/DGP predictive mean of gradient\\ ${\Sigma_{\nabla,\text{GP}}}$,${\Sigma_{\nabla,\text{LGP}}}$,${\Sigma_{\nabla,\text{DGP}}}$ &GP/LGP/DGP predictive covariance of gradient \\
  $\boldsymbol{\theta}$ & set of GP parameters\\
 $\eta$ & nugget term\\
 $\gamma$ & kernel length scale \\
 $\mathcal{D}$ & Dataset \\
 $D_L$ & sharp variation region with threshold \(L\)\\
 $N_{\text{imp}}$ & Number of imputations in DGP emulator \\
 $\text{Jac}(\cdot)$& Jacobian \\
 $\mathbb{E}(\cdot)$ & Expectation \\
 $\text{Var}(\cdot)$ & Variance \\
 $\text{Cov}(\cdot,\cdot)$ & Covariance \\
 $\nabla f(\boldsymbol{x})$ & (nabla) gradient operator on function $f$ at $\boldsymbol{x}$,$\nabla f(\boldsymbol{x})\in\mathbb{R}^d$ \\
 $H_f(\boldsymbol{x})$ & hessian matrix of function $f$ at $\boldsymbol{x}$, $H_f(\boldsymbol{x})\in\mathbb{R}^{d\times d}$\\
 \hline
 \hline
\end{tabular}
}
\end{center}

\section{Derivation of DGP Gradient Distribution Approximation}
\subsection{Proof of Lemma 3.1}
\label{subsec: Proof of Lemma 3.1}
\begin{proof}[Proof of Lemma 3.1]
By the definition of the LGP emulator in Section \ref{subsec: linked gp}, \[y(\boldsymbol{x}_*)=\hat{g}(\boldsymbol{w}(\boldsymbol{x}_*)),\]
where \(\boldsymbol{w}(\boldsymbol{x}_*)\) is the first-layer predictive random vector and \(\hat{g}\) is the second-layer GP. Under LGP construction, conditioning on \((\boldsymbol{W},\boldsymbol{y})\),
\begin{itemize}
    \item \(\hat{g}\) is a GP trained on \((\boldsymbol{W},\boldsymbol{y})\),
    \item the first-layer GPs $\hat{\boldsymbol{f}}$ are independent of $\hat{g}$
\end{itemize}
Hence, by conditional independence and the tower property, 
\[\mu_{\text{LGP}}(\boldsymbol{x}_*)=\mathbb{E}[y(\boldsymbol{x}_*)]=\mathbb{E}_{\boldsymbol{w}(\boldsymbol{x}_*)}\big[\underbrace{\mathbb{E}[\hat{g}(\boldsymbol{w}(\boldsymbol{x}_*))|\boldsymbol{w}(\boldsymbol{x}_*)]}_{\mu_g(\boldsymbol{w}(\boldsymbol{x}_*)):\text{ Posterior Mean given \(\boldsymbol{w}(\boldsymbol{x}_*)\)}}\big].\]
In the following, we first work out \(\nabla_{\boldsymbol{x}_*}\mu_{\text{LGP}}(\boldsymbol{x}_*)\) and then show the equivalence by working from \(\mathbb{E}[\nabla_{\boldsymbol{x}_*}y(\boldsymbol{x}_*)]\).

\paragraph{(1)}\(\nabla_{\boldsymbol{x}_*}\mu_{\text{LGP}}(\boldsymbol{x}_*)\):
Since the first-layer distribution is a diagonal Gaussian:
\[\boldsymbol{w}(\boldsymbol{x}_*)\sim\mathcal{N}(\boldsymbol{\mu}_{\boldsymbol{f}}(\boldsymbol{x}_*),\boldsymbol{\Sigma}_{\boldsymbol{f}}(\boldsymbol{x}_*)),\]
where \(\boldsymbol{\mu}_{\boldsymbol{f}}(\boldsymbol{x}_*)=(\mu_{1}(\boldsymbol{x}_*),\ldots,\mu_{p}(\boldsymbol{x}_*))\) and \(\boldsymbol{\Sigma}_{\boldsymbol{f}}(\boldsymbol{x}_*)=\text{diag}(\sigma^2_1(\boldsymbol{x}_*),\ldots,\sigma^2_p(\boldsymbol{x}_*))\). We can reparameterize the first-layer predictive distribution with a standard normal vector $\boldsymbol{z}\sim\mathcal{N}(\boldsymbol{0},\boldsymbol{I})$ such that 
\[\boldsymbol{w}(\boldsymbol{x}_*)=\boldsymbol{\mu}_{\boldsymbol{f}}(\boldsymbol{x}_*)+\boldsymbol{L}(\boldsymbol{x}_*)\boldsymbol{z},\]
where $\boldsymbol{L}(\boldsymbol{x}_*)=\text{diag}(\sigma_1(\boldsymbol{x}_*),\ldots,\sigma_p(\boldsymbol{x}_*))$. Therefore, 
\(\mu_{\text{LGP}}(\boldsymbol{x}_*)=\mathbb{E}_{\boldsymbol{z}}[\mu_g(\boldsymbol{\mu}_{\boldsymbol{f}}(\boldsymbol{x}_*)+\boldsymbol{L}(\boldsymbol{x}_*)\boldsymbol{z})].\)
Because all the sub-GPs use SE kernels, we have (1) $\mu_g\in C^1(\mathbb{R}^p)$, where $C^1(\mathbb{R}^p)$ denotes the space of continuously differentiable functions on $\mathbb{R}^p$; (2) $\mu_{i},\sigma_{i}\in C^1(\mathbb{R}^d)$ \cite{rasmussen2006gaussian}; and (3) all involved derivatives have finite moments under $\boldsymbol{z}\sim\mathcal{N}(\boldsymbol{0},\boldsymbol{I})$. By dominated convergence theorem (Chapter 2.4 \cite{cohn2013measure}), we can differentiate under the expectation. As $\boldsymbol{z}$ is independent of $\boldsymbol{x}_*$, this gives
\[\nabla_{\boldsymbol{x}_*}\mu_{\text{LGP}}(\boldsymbol{x}_*)=\mathbb{E}_{\boldsymbol{z}}[\nabla_{\boldsymbol{x}_*}\mu_g(\boldsymbol{\mu}_{\boldsymbol{f}}(\boldsymbol{x}_*)+\boldsymbol{L}(\boldsymbol{x}_*)\boldsymbol{z})].\]
Then, \(\nabla_{\boldsymbol{x}_*}\mu_g(\boldsymbol{\mu}_{\boldsymbol{f}}(\boldsymbol{x}_*)+\boldsymbol{L}(\boldsymbol{x}_*)\boldsymbol{z})=\nabla_{\boldsymbol{x}_*}\mu_g(\boldsymbol{w}(\boldsymbol{x}_*))=(\nabla_{\boldsymbol{w}}\mu_g)(\boldsymbol{w}(\boldsymbol{x}_*))\text{Jac}_{\boldsymbol{x}_*}\boldsymbol{w}(\boldsymbol{x}_*)\).
Therefore, putting over expectation gives
\begin{equation}
    \nabla_{\boldsymbol{x}_*}\mu_{\text{LGP}}(\boldsymbol{x}_*)=\mathbb{E}[(\nabla_{\boldsymbol{w}}\mu_g)(\boldsymbol{w})\text{Jac}_{\boldsymbol{x}_*}\boldsymbol{w}(\boldsymbol{x}_*)]\tag{\ding{72}}.
\end{equation}
\paragraph{(2)} \(\mathbb{E}[\nabla_{\boldsymbol{x}_*}y(\boldsymbol{x}_*)]\): For almost every realization of the GPs (with SE kernels, sample paths are almost surely $C^{\infty}$), the global mapping 
\[\boldsymbol{x}_*\mapsto y(\boldsymbol{x}_*)=\hat{g}(\boldsymbol{w}(\boldsymbol{x}_*))\]
is differentiable, and again by the chain rule,
\(
\nabla_{\boldsymbol{x}_*}y(\boldsymbol{x}_*)=(\nabla_{\boldsymbol{w}}\hat{g})(\boldsymbol{w}(\boldsymbol{x}_*))\text{Jac}_{\boldsymbol{x}_*}\boldsymbol{w}(\boldsymbol{x}_*). 
\) Taking expectation and using conditional independence between the first- and second-layer GPs yields:
\begin{equation*}
    \begin{split}
        \mathbb{E}[\nabla_{\boldsymbol{x}_*}y(\boldsymbol{x}_*)]&=\mathbb{E}[(\nabla_{\boldsymbol{w}}\hat{g})(\boldsymbol{w}(\boldsymbol{x}_*))\text{Jac}_{\boldsymbol{x}_*}\boldsymbol{w}(\boldsymbol{x}_*)]\\
        &=\mathbb{E}_{\boldsymbol{w}(\boldsymbol{x}_*)}\big[\mathbb{E}[(\nabla_{\boldsymbol{w}}\hat{g})(\boldsymbol{w}(\boldsymbol{x}_*))|\boldsymbol{w}(\boldsymbol{x}_*)]\text{Jac}_{\boldsymbol{x}_*}\boldsymbol{w}(\boldsymbol{x}_*)\big]
    \end{split}
\end{equation*}
With GP gradient identity as shown in Equation~\eqref{eq:gradient_of_gp}: for the second-layer GP,
\[
\mathbb{E}[\nabla_{\boldsymbol{w}}\hat{g}(\boldsymbol{w})]=\nabla_{\boldsymbol{w}}\mu_g(\boldsymbol{w}),
\]
because differentiation is a linear operator on the GP and the gradient GP has mean equal to the gradient of the mean function. Therefore,
\[\mathbb{E}[\nabla_{\boldsymbol{x}_*}y(\boldsymbol{x}_*)]=\mathbb{E}[(\nabla_{\boldsymbol{w}}\mu_g)(\boldsymbol{w})\text{Jac}_{\boldsymbol{x}_*}\boldsymbol{w}(\boldsymbol{x}_*)].\]
Comparing this with (\ding{72}) gives \(\mathbb{E}[\nabla_{\boldsymbol{x}_*}y(\boldsymbol{x}_*)]=\nabla_{\boldsymbol{x}_*}\mu_{\text{LGP}}(\boldsymbol{x}_*)\). Finally, recall the analytical form of predictive mean, we get 
\[\mathbb{E}[\nabla_{\boldsymbol{x}_*}y(\boldsymbol{x}_*)]=\nabla_{\boldsymbol{x}_*}\mu_{\text{LGP}}(\boldsymbol{x}_*)=\nabla_{\boldsymbol{x}_*} \boldsymbol{I}(\boldsymbol{x}_*)^\top\boldsymbol{R}(\boldsymbol{W})^{-1}\boldsymbol{y}.\]
This completes the proof of Lemma \ref{lemma: exactness by differentiating mean}.
\end{proof}

\subsection{Explicit formula for expectation of LGP gradient under squared exponential kernel} 
The analytical expression of predictive mean of LGP emulator, as Equation \eqref{eq: LGP analytical mean and variance}: \[\mu_{\text{LGP}}(\boldsymbol{x}_*)=\boldsymbol{I}(\boldsymbol{x}_*)^\top\boldsymbol{R}(\boldsymbol{W})^{-1}\boldsymbol{y}.\] As we take the gradient with respect to the $\boldsymbol{x}_*$, the only term involved in $\boldsymbol{x}_*$ is: $\boldsymbol{I}(\boldsymbol{x}_*)\in\mathbb{R}^{n}$: the $i$th entry \[I_i(\boldsymbol{x}_*)=\prod_p\mathbb{E}[k_p\big(w_p(\boldsymbol{x}_*),w_{i,p}\big)]=\prod_p\xi_{ip},\]
where $w_{i,p}$ is the $(i,p)$ entry of $\boldsymbol{W}\in\mathbb{R}^{n\times p}$ and $w_p(\boldsymbol{x}_*)$ is the output of $p$th GP, $\hat{f}_p$.

Given the kernel function $k$ as the squared exponential kernel, $\xi_{ip}$ and $\psi_{ijp}$ have analytical expressions as\footnote{More details can be found in Kyzyurova et al. (2018) \cite{kyzyurova2018coupling} and in Section~3 of Ming and Guillas (2021) \cite{ming2021linked}.}:
\begin{align*}
\xi_{i p} & =\mathbb{E}[k_p(w_p(\boldsymbol{x}_*),w_{i,p})]=\frac{1}{\sqrt{1+2 \sigma_p^2(\boldsymbol{x}_*) / \gamma_p^2}} \exp \left\{-\frac{\left(\mu_p(\boldsymbol{x}_*)-{w_{i,p}}\right)^2}{2 \sigma_p^2(\boldsymbol{x}_*)+\gamma_p^2}\right\}
\end{align*}
where $\gamma_p$ is the length-scale parameter for the $p$th input dimension of the second-layer GP kernel. To compute the $\nabla \mu_{\text{LGP}}(\boldsymbol{x}_*)$, the only unknown term is $\nabla\boldsymbol{I}(\boldsymbol{x}_*)$, which can be analytically computed by applying the multivariate chain rule to the $\boldsymbol{I}(\boldsymbol{x}_*)$:
\begin{equation}
\begin{split}
    &[\nabla \mathbf{I}(\boldsymbol{x}_*)]_{i,j}=\frac{\partial I_i(\boldsymbol{x}_*)}{\partial x_{*,j}}=\sum_{k=1}^p\frac{\partial\xi_{ik}}{\partial x_{*,j}}\prod_{k'\neq k}\xi_{ik'}, \\
    &\text{where }\frac{\partial\xi_{ik}}{\partial x_{*,j}}=\frac{\partial\xi_{ik}}{\partial \mu_k(\boldsymbol{x}_*)}\frac{\partial \mu_k(\boldsymbol{x}_*)}{\partial x_{*,j}}+\frac{\partial\xi_{ik}}{\partial \sigma^2_k(\boldsymbol{x}_*)}\frac{\partial \sigma^2_k(\boldsymbol{x}_*)}{\partial x_{*,j}}
\end{split}
\end{equation}
Then each term can be computed as 
\begin{equation}
\begin{split}
    &\frac{\partial\xi_{ik}}{\partial\mu_k(\boldsymbol{x}_*)}=-\frac{2(\mu_k(\boldsymbol{x}_*)-w_{i,k})}{(2\sigma^2_k(\boldsymbol{x}_*)+\gamma^2_k)\sqrt{1+2\sigma_k^2(\boldsymbol{x}_*)/\gamma_k^2}}\exp{\{-\frac{(\mu_k(\boldsymbol{x}_*)-w_{i,k})^2}{2\sigma^2_k(\boldsymbol{x}_*)+\gamma^2_k}\}}\\
    &\frac{\partial\xi_{ik}}{\partial\sigma^2_k(\boldsymbol{x}_*)}=-\frac{\exp{\{-\frac{(\mu_k(\boldsymbol{x}_*)-w_{i,k})^2}{2\sigma^2_k(\boldsymbol{x}_*)+\gamma_k^2}\}}}{\sqrt{1+2\sigma^2_k(\boldsymbol{x}_*)/\gamma_k^2}}\Big[\frac{1}{\gamma_k^2+2\sigma^2_k(\boldsymbol{x}_*)}-2(\frac{\mu_k(\boldsymbol{x}_*)-w_{i,k}}{2\sigma_k^2(\boldsymbol{x}_*)+\gamma_k^2})^2\Big]\\
    &\frac{\partial\mu_k(\boldsymbol{x}_*)}{\partial x_{*,j}}=\frac{\partial \boldsymbol{r}_k(\boldsymbol{x}_*)}{\partial x_{*,j}}^\top\boldsymbol{R}_k(\boldsymbol{X})^{-1}\boldsymbol{W}_k\\
    &\frac{\partial \sigma^2_k}{\partial x_{*,j}}=c_k^2\big(\frac{\partial k_k(\boldsymbol{x}_*,\boldsymbol{x}_*)}{\partial x_{*,j}}-2\frac{\partial \boldsymbol{r}_k(\boldsymbol{x}_*)}{\partial x_{*,j}}^\top\boldsymbol{R}_k(\boldsymbol{X})^{-1}\boldsymbol{r}_k(\boldsymbol{x}_*)\big).
\end{split}
\end{equation}

\subsection{Derivation of Covariance approximation of LGP gradient - Equation 3.3}
\label{subsec: Derivation of Covariance approximation of LGP gradient - Equation 3.2}
The derivation begins by applying the multivariate chain rule to the composite LGP. The main challenge lies in the nonlinear interaction between the two GP layers, as the gradient of the second-layer, $\hat{g}$, is evaluated at the random output of the first-layer GPs, $\hat{\boldsymbol{f}}$. To make this tractable, we introduce our key approximation using the \textit{delta} method \citep{oehlert1992note}. This step linearizes the problem by evaluating the gradient of $\hat{g}$ at the mean of the first-layer outputs, $\mu_{\boldsymbol{w}_*}$. With the expression linearized, we then apply the law of total variance. This is a crucial step that decomposes the expression into two parts: one capturing the uncertainty from the first-layer GPs ($\hat{\boldsymbol{f}}=(\hat{f}_1,\dots,\hat{f}_p)$) and the other capturing the uncertainty from the second-layer GP ($\hat{g}$). Finally, these two parts are simplified. We expand the terms and apply the assumption that the first-layer GPs are conditionally independent. This step diagonalizes the covariance structure of the Jacobian, leading to the final expression presented in the manuscript. The full mathematical derivation can be shown below: 

\begin{equation}
   \label{eq: full derivation of covaraince approximation}
   \begin{split}
       \Sigma_{\nabla,\text{LGP}} &= \text{Cov}\big[\nabla_{\boldsymbol{x}_*} \hat{g}(\hat{\boldsymbol{f}}(\boldsymbol{x}_*))\big] \\
       &\approx \text{Cov}\big[\nabla_{\boldsymbol{\mu}_{\boldsymbol{w}_*}} \hat{g}(\boldsymbol{\mu}_{\boldsymbol{w}_*})^T \text{Jac}_{\boldsymbol{x}_*}\hat{\boldsymbol{f}}(\boldsymbol{x}_*)\big] \, \text{(Delta method)} \\
       &=\text{Cov}\big[\mathbb{E}[\nabla_{\boldsymbol{\mu}_{\boldsymbol{w}_*}}\hat{g}]^T \text{Jac}_{\boldsymbol{x}_*}\hat{\boldsymbol{f}}\big] \\
       &\quad+\mathbb{E}\big[\text{Jac}_{\boldsymbol{x}_*}\hat{\boldsymbol{f}}^T \text{Cov}(\nabla_{\boldsymbol{\mu}_{\boldsymbol{w}_*}}\hat{g})\text{Jac}_{\boldsymbol{x}_*}\hat{\boldsymbol{f}}\big] \text{(Law of total variance)} \\
       &= \sum_{i=1}^p \left(\mu^{\hat{g}}_{\nabla}\right)_i^2 \Sigma^{\hat{f}_i}_\nabla  + \mathbb{E}[\text{Jac}_{\boldsymbol{x}_*}\hat{\boldsymbol{f}}]^T \Sigma^{\hat{g}}_{\nabla} \mathbb{E}[\text{Jac}_{\boldsymbol{x}_*}\hat{\boldsymbol{f}}] + \sum_{i=1}^p (\Sigma^{\hat{g}}_{\nabla})_{ii} \Sigma^{\hat{f}_i}_\nabla \\
       &= \sum_{i=1}^p \left[ \left(\mu^{\hat{g}}_{\nabla}\right)_i^2 + (\Sigma^{\hat{g}}_{\nabla})_{ii} \right] \Sigma^{\hat{f}_i}_\nabla 
       + \sum_{k=1}^p \sum_{l=1}^p (\Sigma^{\hat{g}}_{\nabla})_{kl} \mu_{\nabla}^{\hat{f}_k} (\mu_{\nabla}^{\hat{f}_l})^T
   \end{split}
\end{equation}
Each term in this sum has a clear interpretation, corresponding to the different ways in which the mean and variance of the two layers interact. 
\section{Proof of proposition 4.1}
\begin{definition}[Noncentral Chi Distribution \citep{lawrence2023moments}]
    A random variable \( Y \) follows a \textit{noncentral chi distribution} with degrees of freedom \( k \) and noncentrality parameter \( \lambda \) if it is defined as:
    \(
    Y = \sqrt{\sum_{i=1}^{k} X_i^2}
    \)
    where \( X_1, X_2, \dots, X_k \) are independent normal random variables such that: $X_i \sim \mathcal{N}(\mu_i, 1)$ for each \( i \), and the noncentrality parameter is given by $\lambda = \sqrt{\sum_{i=1}^{k} \mu_i^2}$. The probability density function and cumulative distribution function are given as
    \begin{equation*}
        \begin{split}
        & f_Y(y; k, \lambda) = e^{-(y^2 + \lambda^2)/2} \left(\frac{y}{\lambda}\right)^{k/2} \lambda \, I_{k/2 -1}(\lambda y), \\
        &F_Y(y;k,\lambda)=1-Q_{\frac{k}{2}}(\lambda,y)
        \end{split}
    \end{equation*}
where \( I_{\nu}(\cdot) \) is the modified Bessel function of the first kind, and $Q_{\frac{k}{2}}(\lambda,y)$ is the Marcum Q-function $Q_M(a,b)$.
\end{definition}

\begin{definition}[Generalized noncentral chi-square distribution \cite{davies1980distribution}]
\label{definition: Generalized noncentral chi-square distribution}

Let \(X_1\)...\(X_k\) be independent \(N(\delta_i,1)\) random variables and let \(
Y_{sq} \;=\; \sum_{i=1}^k a_i X_i^2\), \(a_i > 0.
\)
Then \(Y_{sq}\) is said to follow a \emph{generalized noncentral chi-square distribution}
with weights \(\{a_i\}_{i=1}^k\) and noncentrality parameters
\(\{\delta_i\}_{i=1}^k\).
\(F_{\mathrm{G}\chi^2}(\,\cdot\,; \{a_i\},\{\delta_i\})\) denotes its CDF.
If all \(a_i \equiv 1\), this reduces to the standard noncentral chi-square
distribution, and \(\sqrt{Y_{sq}}\) has the noncentral chi distribution above.
\end{definition}

\begin{proof}[Proof of proposition 4.1]
\label{proof: proposition 5.1}
    Suppose 
    \(
    \nabla f(\boldsymbol{x})
    \;\sim\; \mathcal{N}\bigl(\mu_{\nabla}(\boldsymbol{x}),
                               \Sigma_{\nabla}(\boldsymbol{x})\bigr).
    \)
    As \(\Sigma_{\nabla}(\boldsymbol{x})\) is symmetric positive definite, it has a spectral decomposition
    \[
    \Sigma_\nabla(\boldsymbol x)
    = U(\boldsymbol x)\,\Lambda(\boldsymbol x)\,U(\boldsymbol x)^\top,
    \]
    where $U(\boldsymbol x)$ is orthogonal and
    $\Lambda(\boldsymbol x)=\mathrm{diag}\bigl(\lambda_1(\boldsymbol x),\dots,
    \lambda_d(\boldsymbol x)\bigr)$ with $\lambda_i(\boldsymbol x)>0$ the eigenvalues
    of $\Sigma_\nabla(\boldsymbol x)$. Define the rotated gradient \(
    \tilde G(\boldsymbol x)
    := U(\boldsymbol x)^\top \nabla f(\boldsymbol x).
    \)
    Since orthogonal transforms preserve Gaussianity,
    \[
    \tilde G(\boldsymbol x)\mid\mathcal D
    \;\sim\;
    \mathcal N\bigl(\tilde\mu(\boldsymbol x),\,\Lambda(\boldsymbol x)\bigr),
    \quad
    \tilde\mu(\boldsymbol x):=U(\boldsymbol x)^\top\mu_\nabla(\boldsymbol x).
    \]
    We can write each component as
    \[
    \tilde G_i(\boldsymbol x)
    = \tilde\mu_i(\boldsymbol x) + \sqrt{\lambda_i(\boldsymbol x)}\,Z_i,
    \qquad
    Z_i\sim\mathcal N(0,1)\ \text{independently},\ i=1,\dots,d.
    \]
     Orthogonal transformations preserve the Euclidean norm, so
    \(n_f(\boldsymbol x)= \|\nabla f(\boldsymbol x)\|= \|\tilde G(\boldsymbol x)\|.\) Then, define
    \(
    Y_i
    := \frac{\tilde\mu_i(\boldsymbol x)}{\sqrt{\lambda_i(\boldsymbol x)}}
       + Z_i\), and \(\delta_i
    := \frac{\tilde\mu_i(\boldsymbol x)}{\sqrt{\lambda_i(\boldsymbol x)}},
    \)
    The squared gradient norm can then be expressed as
    \[
    \|\nabla f(\boldsymbol{x})\|^2
     = \sum_{i=1}^d
       \bigl(\tilde\mu_i(\boldsymbol x)
            + \sqrt{\lambda_i(\boldsymbol x)} Z_i\bigr)^2 
    = \sum_{i=1}^d
       \lambda_i(\boldsymbol x)\, Y_i^2.
    \]
    Following Definition \ref{definition: Generalized noncentral chi-square distribution}, 
    $\|\nabla f(\boldsymbol{x})\|^2$ is a generalized noncentral chi-square random variable with weights $w_i = \lambda_i(\boldsymbol x)$ and noncentrality parameters $\delta_i = \tilde\mu_i(\boldsymbol x) / \sqrt{\lambda_i(\boldsymbol x)}$. Let $F_{\mathrm{G}\chi^2}$ denote the corresponding CDF. Then, for any threshold
    $L\ge 0$,
    \[
    p_{\boldsymbol x}
    = p\bigl(\|\nabla f(\boldsymbol{x})\|\ge L\bigr)
    = p\bigl(\|\nabla f(\boldsymbol{x})\|^2\ge L^2\bigr)
    = 1 - F_{\mathrm{G}\chi^2}\!\Bigl(
            L^2;\,
            \boldsymbol{w}(\boldsymbol{x}),\,
            \boldsymbol{\delta}(\boldsymbol{x})
          \Bigr),
    \]
which completes the proof.
\end{proof}


\bibliographystyle{siamplain}
\bibliography{references}

\end{document}